\documentclass[%
 aip, pof,
 amsmath,amssymb,
 reprint,
]{revtex4-1}

\usepackage{graphicx}
\usepackage{dcolumn}
\usepackage{bm}
\usepackage[utf8]{inputenc}
\usepackage[T1]{fontenc}
\usepackage{mathptmx}
\usepackage{etoolbox}
\usepackage[english]{babel}

\usepackage{siunitx}

\usepackage[colorlinks=true, allcolors=blue]{hyperref}

\makeatletter
\expandafter\let\csname l@en\endcsname\l@english
\makeatother

\begin{document}

\title{Physics of the droplet-to-ion transition in electrosprays of highly conducting liquids}

\author{Manel Caballero-P\'erez}
 \email[Electronic mail: ]{manelc@uci.edu}
\affiliation{University of California, Irvine}

\author{Manuel Gamero-Casta\~no}
 \email[Electronic mail: ]{mgameroc@uci.edu}
\affiliation{University of California, Irvine}

\date{\today}

\begin{abstract}
We investigate the physical mechanisms governing the continuous transition from the droplet-dominated to the ion-dominated regime in electrosprays of highly conducting liquids. We characterize electrosprays of four ionic liquids using time-of-flight spectrometry and direct flow rate measurements. In the droplet regime, the jet breakup process exhibits self-similar lognormal mass-to-charge distributions with a constant coefficient of variation. In the mixed and ionic regimes, the average solvation state of the emitted ions decreases with decreasing flow rate, consistent with a shift of the primary ion emission zone toward the cooler cone-jet neck. Modeling ion evaporation from the post-breakup droplet population yields an estimate for the ion solvation energy of $\Delta G_0 \gtrsim 1.9$~eV, a value difficult to reconcile with jet-less ion emission from a Taylor cone tip. Furthermore, we identify two fundamental limits on the performance of highly conducting electrosprays near minimum flow rate: substantial neutral mass losses driven by the evaporation of small droplets, and a dissociation limit imposed by the finite fraction of free ions in the bulk liquid. The dissociation limit yields an analytical expression for the maximum specific impulse of electrospray thrusters, showing excellent agreement with experimental data across multiple propellants and electrospray sources.
\end{abstract}

\maketitle

\section{Introduction}

Electrospraying in the steady cone-jet mode is an electrohydrodynamic phenomenon in which conductive liquids are accelerated through a stationary continuum structure to high speeds. Under strong electric field stress, a conducting liquid meniscus evolves into a Taylor cone from whose apex emerges a thin charged jet that can reach velocities of hundreds of meters per second. This jet eventually becomes unstable, breaking into charged droplets and molecular ions that can be further accelerated electrostatically. This soft ionization technique has enabled breakthroughs in mass spectrometry of biomolecules~\cite{Fenn1989}, controlled nanomaterial fabrication~\cite{He2020}, and satellite micro-propulsion~\cite{Gamero-Castano2001, Lozano2005}.

Liquid conductivity spans many orders of magnitude and is one of the main variables that determines electrospray characteristics. While low to moderate conductivity liquids ($K\sim10^{-10}$--$0.1$ S/m) have been extensively characterized and exhibit isothermal behavior with micrometer to millimeter droplet production \cite{Ganan-Calvo2018}, highly conducting fluids ($K \gtrsim 0.1$~S/m) exhibit distinct physics that is comparatively less explored. These liquids produce nanometric droplets and molecular ions, with some electrosprays emitting purely ionic beams at sufficiently low flow rates \cite{Romero-Sanz2003}. The generation of species with high charge-to-mass ratios makes them especially valuable for applications requiring particle speeds on the order of 1--10~km/s, such as sputtering \cite{Gamero-Castano2009b} and space propulsion \cite{Cisquella-Serra2022, Krejci2016}.

The modeling and characterization of highly conducting liquids presents unique challenges. The jets produced have nanometric radii well below the limits of optical imaging. Ohmic dissipation significantly alters liquid properties throughout the cone-jet geometry \cite{Magnani2024}, while the intense electric fields (of order 1 V/nm) necessitate operation in high vacuum to prevent electrical discharge. By providing a stable source of droplets under extreme electric stress, electrosprays of highly conducting liquids are useful for studying ion field evaporation, a process also central to the physics of field desorption mass spectrometry \cite{Beckey1975}, liquid metal ion sources \cite{Gomer1979}, and atom probe tomography \cite{Kelly2007}.

The flow rate supplied to the electrospray fundamentally controls beam characteristics. At high flow rates, the spray produces droplets with lower charge-to-mass ratios. As flow rate decreases, droplets become smaller and more highly charged, and at sufficiently low flow rates, molecular ions may be emitted. The transition from droplet-dominated to ion-dominated emission in highly conducting liquid electrosprays is accompanied by fundamental changes in the emission mechanisms. As flow rate approaches the minimum stable value, high current densities in the cone-jet transition region cause significant Joule heating, while ion field evaporation from both the cone-jet and droplets becomes increasingly important.

This work extends our recent study demonstrating that smaller emitters stabilize cone-jets at lower flow rates~\cite{Caballero-Perez}, in some cases enabling the emission of a stable beam of molecular ions from a capillary-emitter electrospray. Here, we investigate the fundamental physics governing the droplet-to-ion transition in highly conducting ionic liquids, characterizing the beams and performance of four propellants: 1-ethyl-3-methylimidazolium bis(trifluoromethylsulfonyl)imide (EMI-Im), 1-ethyl-3-methylimidazolium trifluoroacetate (EMI-TFA), 1-butyl-3-methylimidazolium tricyanomethanide (BMI-TCM), and ethylammonium nitrate (EAN). Their physical properties and experimental fitting parameters are listed in Table~\ref{tab:main}.

The paper is organized as follows. Section~\ref{sec:theoretical_background} reviews cone-jet electrospray physics and the theoretical framework for current scaling, jet breakup, and ion evaporation. Section~\ref{sec:experimental} describes our experimental setup and methods. Section~\ref{results} presents results on current-flow rate relationships, beam composition evolution, ion solvation energy estimation, and the dissociation limit. Given our research group's interest in space propulsion applications, we also provide propulsion-relevant analyses in Section~\ref{sec:propulsion}, including neutral propellant losses, analytical predictions for propulsive efficiency, and fundamental limits to electrospray specific impulse. Concluding remarks are presented in Section~\ref{sec:conclusion}.

\begin{table}[hbt!]
\caption{\label{tab:main} Parameters and physical properties for the ionic liquids studied. $\dagger$: parameters at current crossover; $\ddagger$: parameters at $500\times r_c$ from the cone vertex; $*$: unavailable in the literature, estimated from dielectric constants of similar ionic liquids. Physical properties $K$, $\gamma$, $\mu$, $\rho$ are given at 295~K.}
\centering
\begin{tabular}{llcccc}
    \hline
    \hline
    Parameter & Units & EMI Im & EMI TFA & BMI TCM & EAN \\
    \hline
    $b_1^\dagger$ & K$\cdot$s/kg & 0.526 & 0.308 & 0.213 & 0.479 \\
    $b_2^\dagger$ & - & -0.400 & -0.420 & -0.259 & -0.243 \\
    $b_3^\dagger$ & K & -7.63 & -8.14 & -28.2 & -39.2 \\
    \hline
    $b_1^\ddagger$ & K$\cdot$s/kg & 0.0104 & 0.00599 & 0.130 & 1.27 \\
    $b_2^\ddagger$ & - & -0.390 & -0.414 & -0.292 & -0.223 \\
    $b_3^\ddagger$ & K & -3.61 & -5.43 & -10.2 & -52.9 \\
    \hline
    $P_\gamma$ & \SI{}{\micro \newton / \kelvin} & -51.45 & -51.45 & -71.66 & -76.43 \\
    $S_\gamma$ & \SI{}{\milli \newton} & 51.15   & 47.79   & 71.26   & 72.39 \\
    $P_\rho$  & kg/(m$^3$\,K) & -1.00    & -0.796  & -0.672   & -0.603  \\
    $S_\rho$   & kg/m$^3$ & 1816      & 1529    & 1247      & 1392 \\
    $Y_\mu$ & \SI{}{\milli \pascal \second} & 0.19058 & 0.43386 & 0.20639 & 8.77$\times10^{4}$ \\
    $B_\mu$ & K & 738.51 & 507.64 & 572.92 & 3963.6 \\
    $T_\mu$ & K & -154.48 & -180.28 & -181.44 & 76.061 \\
    $Y_K$ & S/m & 58.0 & 96.4 & 12.24 & 80.14 \\
    $B_K$ & K & -554.0 & -597.5 & -142.2 & -389.7 \\
    $T_K$ & K & -165.0 & -167.4 & -242.7 & -188.6 \\
    \hline
    $K$ & S/m & 0.82 & 0.89 & 0.81 & 2.06 \\
    $\gamma$ & mN/m & 36.0 & 32.6 & 50.1 & 49.8 \\
    $\mu$ & mPa$\cdot$s & 36.5 & 36.2 & 32.04 & 38.2 \\
    $\rho$ & g/cm$^3$ & 1.52 & 1.29 & 1.05 & 1.21 \\
    $\varepsilon$ & - & 13.8 & 14$^{*}$ & 14$^{*}$ & 29 \\
    $M^+$ & Da & 111.2 & 111.2 & 139.2 & 46.1 \\
    $M^-$ & Da & 280.1 & 113.0 & 90.1 & 62.0 \\
    $Re$ & - & 0.0076 & 0.0066 & 0.0096 & 0.0062 \\
    \hline
    $h_1$ & - & 1.72 & 1.82 & 2.00 & 2.22 \\
    $h_2$ & - & 0.310 & 0.305 & 0.293 & 0.293 \\
    \hline
    $\psi$ & - & 2.6 & 2.3 & 2.7 & 2.15 \\
    $u$ & - & 0.706 & 0.820 & 0.832 & 0.872 \\
    $z$ & - & N/A & N/A & 3.5 & 1.55 \\
    $\nu$ & - & N/A & N/A & 0.130 & 0.095 \\
    \hline
    $\Delta H_v$ & kJ/mol & 119.9 & 120.5 & N/A & 100.6 \\
    $w$ & - & 27.32 & 27.64 & N/A & 38.1 \\
    \hline
    \hline
\end{tabular}
\end{table}

\section{Theoretical background}\label{sec:theoretical_background}

\subsection{Cone-jet electrospray physics}
A cone-jet comprises four distinct regions: the upstream Taylor cone, the cone-jet transition region, the jet, and the jet breakup zone. In the cone region, charge transport occurs via ohmic conduction with negligible voltage drop, whereas surface convection of charge dominates in the jet \cite{Gamero-Castano2019}. In the transition region, the transport mechanism evolves from conduction to convection, and electric fields reach local maxima. This transition region sets the emitted current and likely influences the minimum flow rate at which electrosprays can operate stably \cite{Ponce-Torres2018, Gamero-Castano2019a}.

For low to medium conductivity liquids, the isothermal cone-jet problem is characterized by three dimensionless numbers: the dielectric constant $\varepsilon$, the dimensionless flow rate $\Pi$, and the electrohydrodynamic Reynolds number $Re$:
\begin{equation}
   \Pi = \frac{\rho K Q }{\gamma \varepsilon_0}, \qquad  Re = \left(\frac{\gamma^2 \rho \varepsilon_0}{\mu^3 K}\right)^{1/3}
    \label{eq:Pi}
\end{equation}

Here, $\rho$ is the density, $Q$ is the volumetric flow rate, $\gamma$ is the surface tension, $\varepsilon_0$ is the permittivity of vacuum, and $\mu$ is the viscosity. The highly conducting ionic liquids studied here exhibit Reynolds numbers below 0.01, substantially lower than the 0.1--10 range typical of organic solvents extensively studied in electrospray literature \cite{Ganan-Calvo2009}. This leads to self-heating of the fluid that begins in the transition region and intensifies downstream. As a result, the fluid's physical properties can change drastically throughout the cone-jet and the non-isothermal problem requires additional parameters to be fully characterized. Throughout this text, we will use the three dimensionless numbers of the isothermal problem along with modifications to capture non-isothermal behavior. All parameters and physical properties are referred to their values at the upstream temperature of 295~K, unless otherwise specified.

The cone-jet operates stably within a specific voltage-flow rate parameter space \cite{Cloupeau1989}. The minimum flow rate is particularly significant for applications, as decreasing flow rate produces narrower, more monodisperse beams with higher charge-to-mass ratios. For electrospray propulsion, operation near minimum flow rate is especially attractive as specific impulse increases with charge-to-mass ratio \cite{Gamero-Castano2001}. The minimum flow rate mechanism has been attributed to instabilities in the transition region where electric stress dominates and is balanced by viscous stresses \cite{Ganan-Calvo2013, Higuera2017, Gamero-Castano2019a}, yielding:
\begin{equation}
        \Pi_{\text{min}} \sim Re^{-1}
        \label{eq:minflow_visc}
\end{equation}

For highly conducting liquids, it has been shown that emitter geometry strongly affects the minimum flow rate. For example, the same liquid electrosprayed from a \SI{15}{\micro\meter} diameter tip exhibits minimum flow rates an order of magnitude lower than from a \SI{50}{\micro\meter} tip \cite{Caballero-Perez}. The stabilization mechanism remains unclear. The emitted current $I$ follows:
\begin{equation}
    \frac{I}{I_0} \cong \psi \Pi^{1/2}
    \label{eq:curr_scaling}
\end{equation}
where $I_0 = \sqrt{\varepsilon_0\gamma^2/\rho}$ is the current scale, and $\psi \approx 2.5$ for most liquids. Jet features scale with characteristic length $r_c$ \cite{Perez-Lorenzo2022, Magnani2025a}:
\begin{equation}
    \frac{r_c}{r_0} = \Pi^{1/2}
    \label{eq:r_c}
\end{equation}
where $r_0 = \left( (\varepsilon_0^2 \gamma)/(\rho K^2) \right)^{1/3}$ is the length scale. An alternative scaling $r_l = r_0 \Pi^{5/9}$ has recently been proposed that may better match experimental data and has a stronger theoretical basis \cite{Magnani2025a, Perez-Lorenzo2022}.

As mentioned earlier, self-heating from ohmic and viscous dissipation is critical in electrosprays of highly conducting liquids. Dissipation occurs primarily in the transition region and continues along the jet, increasing liquid temperature and causing irreversible voltage drop \cite{Gamero-Castano2010, Magnani2024, Magnani2025}. Numerical studies yield temperature rise well-fitted by \cite{Magnani2024}:
\begin{equation}
    \Delta T = b_1 \dot{m}^{-b_2} + b_3
    \label{eq:tempincrease}
\end{equation}

where $\dot{m}$ is the mass flow rate and the parameters $b_1$, $b_2$, and $b_3$ depend on the position within the cone-jet. Temperature changes alter liquid properties as follows \cite{Magnani2024}:
\begin{align}
    \gamma(T) &= P_\gamma T + S_\gamma\\
    \rho(T) &= P_\rho T + S_\rho\\
    \mu(T) &= Y_\mu \exp\left(\frac{B_\mu}{T+T_\mu}\right)\\
    K(T) &= Y_K \exp\left(\frac{B_K}{T+T_K}\right) \\
     p(T) &= \exp\left(\frac{-\Delta H_{\text{vap}}}{RT} + w\right)\exp \left( \frac{2\gamma M_p}{\bar{r} \rho R T} \right)
    \label{eq:clausius}
\end{align}

where $p$ is the vapor pressure, $\Delta H_\text{vap}$ is the enthalpy of vaporization, $R$ is the gas constant, $M_p=M^++M^-$ is the molar mass of a cation-anion pair, and $\bar{r}$ is the mean radius of curvature of the liquid surface. The fitting parameters $w$, $P_\times$, $S_\times$, $Y_\times$, $B_\times$, and $T_\times$ are given in Table \ref{tab:main}. In the vapor pressure expression, the first exponential is the Clausius-Clapeyron relation and the second is the Kelvin correction for surface curvature \cite{Mitropoulos2008}. Highly conducting liquid electrosprays generate electric fields of order 1~V/nm, sufficient to induce molecular ion evaporation. Local field maxima occur at the cone-jet neck (approximately at the point of maximum curvature), during jet breakup, and at droplet surfaces. A numerical model determined that the maximum electric field at the cone-jet neck scales as \cite{Magnani2024}:

\begin{equation}
    \frac{{E}_m}{E_0}  = h_1 \Pi^{h_2}
    \label{eq:e_conejet}
\end{equation}

where $E_0 = \gamma^{5/6} \varepsilon_0^{-1/3} \rho^{-1/3} K^{-1/6} Q^{-1/2}$ is the electric field scale, and $h_1$ and $h_2$ are fitting parameters listed in Table \ref{tab:main}. It can be readily found that this electric field is of similar magnitude to that at the droplets.

\begin{figure*}[t!]\centering

    \includegraphics[width=\linewidth]{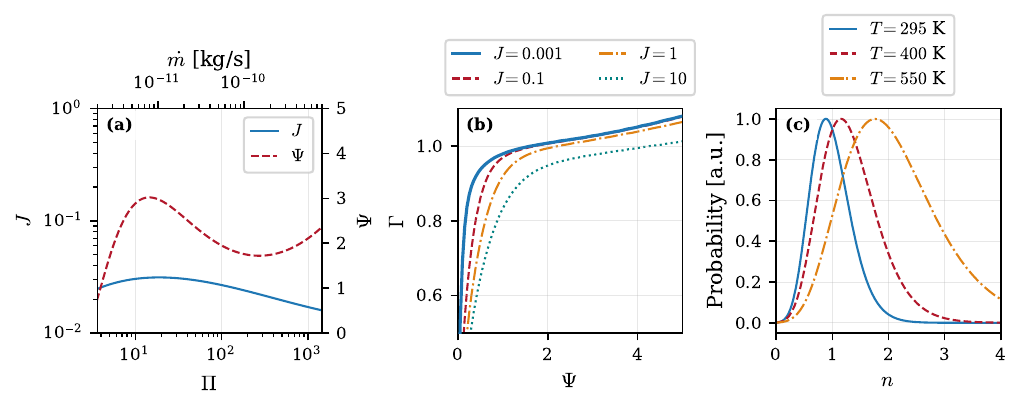}
    \caption{(a) $J$ parameter and Taylor number $\Psi$ as functions of flow
rate for BMI-TCM. (b) Ratio of the critical droplet charge to its Rayleigh charge $\Gamma$ as a function of the Taylor number $\Psi$. (c) Relative probability density as a function of the ion solvation state $n$.}
    \label{fig:j_psi}
\end{figure*}

\subsection{Jet breakup}\label{sec:jet_breakup}

The cone-jet scaling laws presented thus far do not account for ion field evaporation from the cone-jet. For the liquids we study, we observe a large fraction of the electrospray current ejected as ions, and we argue later that a significant portion likely originates from the cone-jet rather than the breakup region and droplets. For that reason, the expressions of this section may have limited accuracy in regimes where evaporation becomes significant. 

The jet breaks up at radius $r_j$ and distance $l_j$ from the neck apex. Several experimental studies calculated $r_j$ by determining the jet velocity at the breakup point $v_j$, and found it to be proportional to $r_c$ for a wide range of flow rates ($\Pi=600-8000$):

\begin{align}
    \frac{r_j}{r_c} &= \tilde{r}_j \approx \text{constant} \label{eq:rj}
\end{align}

The jet radii for the liquids EMI-FAP, EMI-Im, BMI-Im, BMI-DCA, BMI-BF$_4$\footnote{BMI: 1-butyl-3-methylimidazolium, DCA: dicyanamide, BF$_4$: tetrafluoroborate, FAP: tris(pentafluoroethyl)trifluorophosphate} were in the range $\tilde{r}_j=0.23-0.31$, with an average of $\tilde{r}_j\approx0.28$ \cite{Miller2021, Perez-Lorenzo2022, Gamero-Castano2021}.
Ohmic and viscous dissipation lead to irreversible voltage losses throughout the jet. At the breakup point, the experimental voltage loss for EMI-Im cone-jets was found to be \cite{Magnani2024, Gamero-Castano2021}:

\begin{equation}
    \frac{\Delta V}{V_0} \approx 12.91 + 9.787 \sqrt{\Pi} \label{eq:delta_v}
\end{equation}

where $V_0=\pi^{-1} \left( \gamma^4 / (\varepsilon_0 \rho K^2) \right)^{1/6}$ is the electric potential scale. Reference \citenum{Magnani2024} matched for EMI-Im the experimental voltage at the jet breakup with the voltage obtained from cone-jet numerical models, obtaining an approximate jet length:

\begin{equation}
    \frac{l_j}{r_c} = \tilde{l}_j \approx 48.03 + 8.133 \sqrt{\Pi} \label{eq:lj}
\end{equation}

Two dimensionless numbers characterize the breakup: the $J$ parameter, which compares inertial to viscous stresses, and the Taylor number $\Psi$, which quantifies jet electrification:

\begin{align}
    J &= \frac{\rho \gamma r_j}{\mu^2} \approx \tilde{r}_j \Pi^{1/2} Re^2 \label{eq:jnumber}\\
    \Psi &= \frac{\sigma^2 r_j}{\varepsilon_0 \gamma}\label{eq:taylornumber}
\end{align}

where $\sigma = \rho r_j/(2 \xi_j)$ is the jet surface charge and $\xi_j$ is the jet mass-to-charge ratio, or jet mass/charge for short. When $\Psi=2$, electric and surface stresses balance. Substituting Eq.~\eqref{eq:minflow_visc} yields $\Psi \sim 0.1 Re^{-1/2}$ and $J\sim 0.4 Re^{3/2}$. Thus, highly conducting ionic liquids near minimum flow rate are characterized by highly electrified, viscous breakups. With viscous effects dominating, breakup evolves on the viscous timescale $t_\mu=\mu r_j/\gamma$ \cite{Misra2022}. Since the electrical relaxation time $t_k=\varepsilon \varepsilon_0/K$ is much smaller than $t_\mu$ for highly conducting liquids, charge remains relaxed to the surface as the jet evolves, yielding nearly equipotential breakup \cite{Gamero-Castano2021}. Large electrification numbers are typically associated with whipping jets, characterized by chaotic lateral oscillations, whereas lower values lead to varicose breakups \cite{Yang2014}.

Figure~\ref{fig:j_psi}(a) plots Eqs. \eqref{eq:jnumber} and \eqref{eq:taylornumber} for BMI-TCM, using the average jet radius from other liquids $\tilde{r}_j=0.28$ and physical properties at the jet. These values are approximate: we lack the breakup radius for BMI-TCM, and more importantly, we assume no current loss before breakup, a simplification valid only at high flow rates. At $\Pi \lesssim 100$, a significant fraction of the total current evaporates from the cone-jet neck (see Section \ref{sec:dg}). Defining the ratio of the current remaining at the jet to the total current as $\varphi = \xi_t /\xi_j$, where $\xi_t = \dot{m}/I$, and noting that the Taylor number is proportional to $\varphi^2$, we expect $\Psi$ at $\Pi \lesssim 100$ to be lower than predicted in Fig.~\ref{fig:j_psi}(a).

As the jet begins to destabilize, perturbations of different wavelengths $\lambda$ grow at different rates. A critical wavelength $\lambda^*$ with maximum growth rate dominates the breakup, determining pinch-off spacing and consequently the radius $r^*$ of critical (mean) droplets. These droplets approximately inherit the jet's mass/charge, while the inherent randomness produces a distribution in size and charge centered around critical values. The critical droplets with charge $q^*$ and radius $r^*$ can be seen to precede from an unperturbed jet section of length $\lambda^*$. From conservation of charge and mass, we obtain:

\begin{align}
    \lambda^* &= \frac{4 {(r^*)}^3}{3 r_j^2}
    \label{eq:thetabreakup} \\
    q^* &= \sigma 2 \pi r_j \lambda^*
\end{align}

These relations determine the droplet-to-jet radius ratio $\vartheta$ and their charge relative to the Rayleigh limit $\Gamma$:

\begin{align}
    \vartheta &= \frac{r^*}{r_j}= \left(\frac{3\pi}{2x^*}\right)^{1/3}
    \label{eq:vartheta_linearstability} \\
    \Gamma &= \frac{q^*}{q_R(r^*)}=\frac{\vartheta^{3/2}\Psi^{1/2}}{3}
\end{align}

where $x^*=2\pi r_j/\lambda^*$ is the critical wavenumber and the Rayleigh limit is given by:

\begin{equation}
    q_{R} = 8 \pi \sqrt{\gamma \varepsilon_0 {r}^3}
    \label{eq:rayleigh}
\end{equation}

From the previous relations, the critical droplet radius and electric field are:

\begin{align}
    r^* &= \left(\frac{36  \Gamma ^2\xi_t^2  \gamma \varepsilon_0}{\rho^2 \varphi^2} \right)^{1/3} \label{eq:rstar} \\
    E^* &=\frac{q^*}{4\pi\varepsilon_0(r^*)^2}= \left(\frac{4 \rho \gamma \varphi \Gamma^2}{3 \xi_t \varepsilon_0^2}\right)^{1/3}
    \label{eq:estar}
\end{align}

The main ambiguities in these expressions stem from $\varphi$ and $\Gamma$. For the former, we will argue in Section \ref{sec:dg} that negligible ion evaporation occurs from the cone-jet neck for $\Pi\gtrsim100$, hence $\varphi \cong 1$ in that case. A study determined $\Gamma$ experimentally for EMI-Im and found it to be within approximately 0.5 and 1 \cite{Gamero-Castano2009}. The critical wavenumber $x^*$ and hence $\Gamma$ can be approximated from a linear instability analysis, which will be presented later in this section. However, this is only valid for axisymmetric, varicose breakups, which likely does not apply to our jets due to their high electrification.

The equipotential breakup condition constrains a family of droplets of a given mass/charge to a common characteristic potential \cite{Gamero-Castano2021}

\begin{equation}
    \phi_c = \frac{q}{4\pi \varepsilon_0 r},
    \label{eq:constraint}
\end{equation}

which may be approximated by the potential of the most likely droplet, $\phi_c \approx \rho (r^*)^2/(3\varepsilon_0 \xi_j)$. Substituting this expression into Eq.~\eqref{eq:constraint} yields the radius and field distributions as a function of the mass/charge of a family of droplets

\begin{equation}
    r(\xi) = r^*\left(\frac{\xi}{\xi_j}\right)^{1/2}, \qquad E(\xi) = E^* \left(\frac{\xi_j}{\xi}\right)^{1/2}.
    \label{eq:r_E_distr}
\end{equation}

For an axisymmetric, varicose, and equipotential breakup, the growth rate $\omega$ of a perturbation with wavenumber $x=2\pi r_j/\lambda$ is obtained by eliminating $y$ from \cite{Gamero-Castano2002}:

\begin{equation}
     \omega = y^2 - x^2
\end{equation}

\begin{align}
f(x,y,J,\Psi) = \notag \\
2x^2(x^2 + y^2) \frac{\text{I}_1'(x)}{\text{I}_0(x)} \left[ 1 - \frac{2xy \, \text{I}_1(x) \text{I}_1(y)}{x^2 + y^2 \, \text{I}_1'(x) \text{I}_1(y)} \right] - (x^4 - y^4) \notag \\
- J \left\{ x(1 - x^2) \frac{\text{I}_1(x)}{\text{I}_0(x)} - \Psi \frac{x \text{I}_1(x)}{\text{I}_0(x)} \left[ 1 + \frac{x \text{K}_0'(x)}{\text{K}_0(x)} \right] \right\}
\end{align}

where $\text{I}_\alpha$ and $\text{K}_\alpha$ are modified Bessel functions of the first and second kind, with derivatives marked by primes. The critical wavenumber $x^*$ is the one that maximizes $\omega$. Reference \citenum{Gamero-Castano2021} shows that $\vartheta$ collapses to a single function of $\Psi$ for $J\lesssim0.1$ (the high-viscosity limit relevant to our liquids), except at very low Taylor numbers. Figure~\ref{fig:j_psi}(b) plots $\Gamma$ versus $\Psi$ for several $J$.

In the high-viscosity limit ($J\ll1$), Taylor numbers above $\Psi\cong1.65$ produce droplets charged above the Rayleigh limit, making them unstable. Their fate depends on the droplet's initial size and charge. Droplets at or above the Rayleigh limit with radii above 70~nm undergo Rayleigh fission, deforming and extruding thin jets that eject 15--25\% of their charge with minimal mass loss \cite{Schweizer1971, Abbas1967, Richardson1997, Singh2021}.

For radii between 23~nm and 50~nm, EMI-Im droplets charged 4\% above the Rayleigh limit evaporate ions from the intense electric fields at these thin jets, shedding 24\% of the initial parent charge as ions and 16\% in daughter droplets. Below 23~nm, ion evaporation suppresses jet formation; instead, the parent droplet deforms and elongates, evaporating ions from its tips \cite{Misra2023}. We will refer to this phenomenon as Rayleigh ion evaporation. Droplets between 8~nm and 23~nm shed 18--24\% of their charge as ions, while those below 8~nm lose an increasing fraction.

For reference, 8~nm, 23~nm, and 50~nm correspond to the critical droplet radii at $\Pi\cong(\psi/6)^2(r^*/r_0)^3=29$, $688$, and $7063$ for EMI-Im, respectively. Nearly all the experimental data points in our study fall within this range.

The jet breakup and the possible Rayleigh fission and evaporation are not sequential but intertwined, occurring simultaneously within the viscous timescale $t_\mu$~\cite{Misra2022, Misra2023}, as observed in high-speed imagery~\cite{Yang2014}. This timescale is orders of magnitude shorter than the droplet residence time in the acceleration region:

\begin{equation}
t^* \sim \ell \left(\frac{\xi_j}{2V}\right)^{1/2},
\label{eq:t_star}
\end{equation}

where $\ell$ is the emitter-extractor distance and $V$ the acceleration voltage. Consequently, we can treat evaporation during breakup and during flight in the acceleration region as sequential, independent events. Ions evaporating during or shortly after breakup should thus have retarding potentials close to the jet breakup potential.

Previous tandem retarding potential and time-of-flight (TOF) studies of highly conducting ionic liquids \cite{Gamero-Castano2021, Perez-Lorenzo2022, Gamero-Castano2008, Miller2021, Lyne2023} provide insights into the location of ion evaporation. At the flow rates they characterized ($\Pi = 600 - 3600$), ions consistently carry approximately 20\% of the current, with narrow retarding potential distributions centered slightly below the jet breakup potential. This pattern indicates evaporation occurring shortly after breakup rather than continuous evaporation throughout the acceleration region or from the cone-jet.

Below $\Pi \approx 450$, however, a recent study shows that ion current increases with decreasing flow rate \cite{Caballero-Perez}. The origin of these additional ions, whether from the breakup, cone-jet neck, or droplets in flight, remains uncertain.

\subsection{Ion evaporation}

Ions in liquid media, such as solutions or ionic liquids, are bound by ion-dipole and Coulomb interactions. While these forces allow ions to move freely within the bulk, they impose a significant energy barrier to evaporation. Iribarne and Thomson \cite{Iribarne1976, Thomson1979} proposed a rate law for the evaporation of ions from a liquid, describing the ion flux $\Phi$ from a liquid surface with charge density $\sigma$ as:

\begin{equation}
    \Phi = \sigma \, \frac{k_B T}{h} \exp \left( - \frac{\Delta G_0 -G(E,\kappa)}{k_BT} \right)
    \label{eq:iribarne}
\end{equation}

where $k_B$ and $h$ are the Boltzmann and Planck constants, respectively. The ion solvation energy $\Delta G_0$ represents the energy barrier in the absence of an external electric field. For ionic liquids, this value has not been determined experimentally but is typically assumed to lie between 1 and 2\,eV. If ions are evaporated from a surface with non-zero electric field, the effective energy barrier is reduced due to the electric repulsion between the ion and the surface of the same polarity by a quantity $G$, dependent on the surface's electric field and curvature, $E$ and $\kappa$, respectively. For emission from droplets, modeled as charged dielectric spheres of radius $r$ and electric field $E$, $G$ is \cite{Magnani2022}:

\begin{align}
G(E,\kappa) &= \zeta(\kappa) G_p(E) \label{eq:G}\\
\zeta(\kappa)&=\sqrt{\frac{\varepsilon+1}{\varepsilon-1}}\left[\frac{1}{2(1 + \kappa)} + \kappa \sum_{n=1}^{\infty} \frac{\varepsilon - 1}{\varepsilon + 1 + \frac{1}{n}} \left( \frac{1}{1 + \kappa} \right)^{2n+2}\right] \\
G_p(E) &= \left(\frac{e^3 E (\varepsilon -1)}{4\pi \varepsilon_0 (\varepsilon+1)}\right)^{1/2}
    \label{eq:delta_gep}
\end{align}

where $\kappa=\left(e/(16 \pi \varepsilon_0 E r^2)\right)^{1/2}$ is the dimensionless droplet curvature. As the droplet radius increases ($\kappa \to 0$), the geometric factor $\zeta(\kappa)$ approaches unity, recovering the solution for a charged dielectric plane, $G_p(E)$.

Evaporated ions may carry neutral ion pairs with them. The solvation state $n$ denotes the number of attached neutral pairs, so that a singly charged cluster is written as A$^+$[AB]$_n$: $n=0$ corresponds to a bare monomer ion, $n=1$ to a dimer, and so on.

The ion solvation energy $\Delta G_0$ can be estimated using the Born model. Treating an ion cluster of charge $q$ as a sphere of radius $a$ escaping from a parent droplet of radius $r$, the total energy barrier comprises surface and electrostatic components: \cite{Labowsky2000}

\begin{equation}
    \Delta G = 4 \pi \gamma a^2 + \frac{q^2}{8\pi \varepsilon_0}\left(\frac{1}{a} - \frac{2}{3r}\right)\left(1 -\frac{1}{\varepsilon}\right)
    \label{eq:dg0_1}
\end{equation}

The energy barrier is minimized for a single elementary charge $e$ and at a characteristic cluster radius $a_c$:

\begin{equation}
    a_c =  \frac{1}{4} \left(\frac{e^2 (1-1/\varepsilon)}{\pi^2 \gamma \varepsilon_0}\right)^{1/3}
    \label{eq:radius_cluster}
\end{equation}

which, substituted back into Eq.~\eqref{eq:dg0_1}, yields:

\begin{equation}
     \Delta G_{0} = \underbrace{\left(\frac{\gamma e^4 (1-1/\varepsilon)^2}{\pi \varepsilon_0^2 }\right)^{1/3}}_{\Delta G_{\text{Born}}} \underbrace{- \frac{e^2}{12\pi \varepsilon_0 r }\left(1 -\frac{1}{\varepsilon}\right)}_{\text{curvature correction}}
    \label{eq:dg_born}
\end{equation}

For singly charged BMI-TCM ions, the Born model predicts $\Delta G_{\text{Born}} = 2.2$\,eV, with a relatively minor curvature correction for typical droplet sizes ($\approx 0.1$\,eV for $r=10$\,nm). To evaluate the energy barrier at varying solvation states, we estimate the cluster size using van der Waals volumes. The volumes for [BMI]$^+$ and [TCM]$^-$ are $v^+ = 148.5$\,\AA$^3$ and $v^- = 83.1$\,\AA$^3$, respectively \cite{Yu2021}. An ion cluster of solvation state $n$ (A$^+$[AB]$_n$) has an approximate radius $a(n) = \left(3v/4\pi\right)^{1/3}$, where $v = v^+ + n(v^+ + v^-)$.

The evaporated flux for each cluster size scales with the Boltzmann factor $\exp\left( - \Delta G(n)/k_B T \right)$. Figure~\ref{fig:j_psi}(c) compares this distribution for the BMI-TCM at temperatures of 295~K, 400~K, and 550~K. These temperatures correspond to estimated jet conditions at asymptotically high flow rates (no self-heating), $\Pi = 145$, and $\Pi=7.3$, respectively. The shift arises from the decrease in liquid surface tension: 0.050, 0.043, and 0.032~N/m at 295~K, 400~K, and 550~K, respectively. The Born model predicts an energetic minimum between $n=0$ and $n=1$ at room temperature, shifting to between $n=1$ and $n=2$ at higher temperatures. For smaller molecules such as EAN, the characteristic cluster radius corresponds to higher solvation states, consistent with experimental observations \cite{Caballero-Perez}. However, the Born model incompletely captures the ion emission characteristics observed from large-molecule ionic liquids. Electrosprays of EMI-Im and EMI-BF$_4$ emit comparable proportions of monomers ($n=0$) and dimers ($n=1$), yet Fig.~\ref{fig:j_psi}(c) predicts negligible monomer flux relative to dimers. This discrepancy likely stems from the model's assumption that cluster charge delocalizes over the entire spherical surface, an oversimplification for low-solvation ion clusters and bare molecular ions.

\section{Experimental}\label{sec:experimental}

Our experimental setup and TOF signal acquisition procedures are described in detail in our recent study\cite{Caballero-Perez}. Briefly, the electrospray source operates in a vacuum chamber maintained at $10^{-3}$~\SI{}{\pascal}. The source consists of a \SI{360}{\micro\meter} outer diameter fused silica capillary tapered at the tip. The electrically conductive emitter tip connects to a high-voltage power supply through a nano-ammeter and high-voltage switch. The voltage employed varied from 700~V to 1500~V, depending on the liquid, emitter, and flow rate. The emitter capillary is connected to a flow meter capillary whose inner diameter is comparatively much larger. The end of the flow meter is immersed in an ionic liquid vial placed within a hermetic bottle filled with nitrogen. The pressure differential between bottle and chamber drives propellant flow, with pressure controlled through a manifold connected to vacuum pump and nitrogen cylinder via solenoid valves.

A grounded extractor electrode faces the emitter at approximately 1~mm from its tip, creating a strong electric field. The emitter voltage is adjusted to form a stable cone-jet electrospray. Only positive emitter voltages were used in this study. The beam exits through a \SI{1}{\milli\meter} aperture in the extractor center and traverses a field-free region before reaching a planar collector. The collector is grounded through a fast-response electrometer connected to an oscilloscope. A stainless steel grid held at \SI{-10}{\volt} is positioned before the collector to prevent secondary electron escape. Data acquisition devices (DAQs) are used to record the reservoir pressure, the emitter current, the extractor current, and the emitter voltage. The oscilloscope records collector current during TOF measurements, while DAQs continuously monitor it otherwise. A Python script controls all DAQs, solenoid valves, and oscilloscope. Experiments were conducted at a controlled laboratory temperature of \SI{22\pm1}{\celsius}.

Each ionic liquid underwent 24-hour vacuum degassing. Mass flow rate was measured directly using the flow meter capillary connected in series with the emitter. During flow measurements, the flow meter line was lifted from the ionic liquid vial and exposed to atmospheric pressure, creating a visible air-liquid interface. By measuring the time $\Delta t$ for the interface to traverse distance $\Delta x$, the hydraulic resistance $R_H$ of the emitter capillary is determined:

\begin{equation}
    R_H = \frac{P_{\text{atm}}}{A_{\text{flow}} \Delta x / \Delta t}
    \label{eq:hyd_res}
\end{equation}

where $P_{\text{atm}}$ is atmospheric pressure and $A_{\text{flow}}$ is the flow meter capillary cross-sectional area. Since the flow meter's hydraulic resistance is negligible compared to the emitter's, interface position does not affect total resistance. This procedure enables flow rate calculation as $\dot{m} = \rho P/R_H$ at any pressure $P$. The relative uncertainty in total mass flow rate was approximately 10\% for all measurements~\cite{Caballero-Perez}.

TOF measurements were performed by periodically shorting the emitter voltage to ground and subsequently reapplying it at a rate of $0.1-1$\,Hz. The resulting collector current constitutes the TOF signal. Baseline noise signals, recorded before introducing ionic liquid to the emitter, were subtracted from TOF data. The normalized TOF signal is the complement of the cumulative distribution function (cdf) of the current-weighted TOF distribution, $1-F_\tau(\tau)$, and the negative of its derivative is the probability density function (pdf) $f_\tau (\tau)$. The TOF $\tau$ and mass/charge $\xi$ are related by:

\begin{equation}
    \xi = \left(\frac{\tau}{L}\right)^2 2V
    \label{eq:map}
\end{equation}

The current-weighted mass/charge pdf is then obtained as $f(\xi) = (d\tau/d\xi)\, f_\tau(\tau(\xi))$. This mapping neglects beam angular and retarding potential distributions; associated errors are discussed in Appendix~\ref{appendix:mapping}.

\section{Results and discussion}\label{results}

\begin{figure*}[]\centering

    \includegraphics[width=\linewidth]{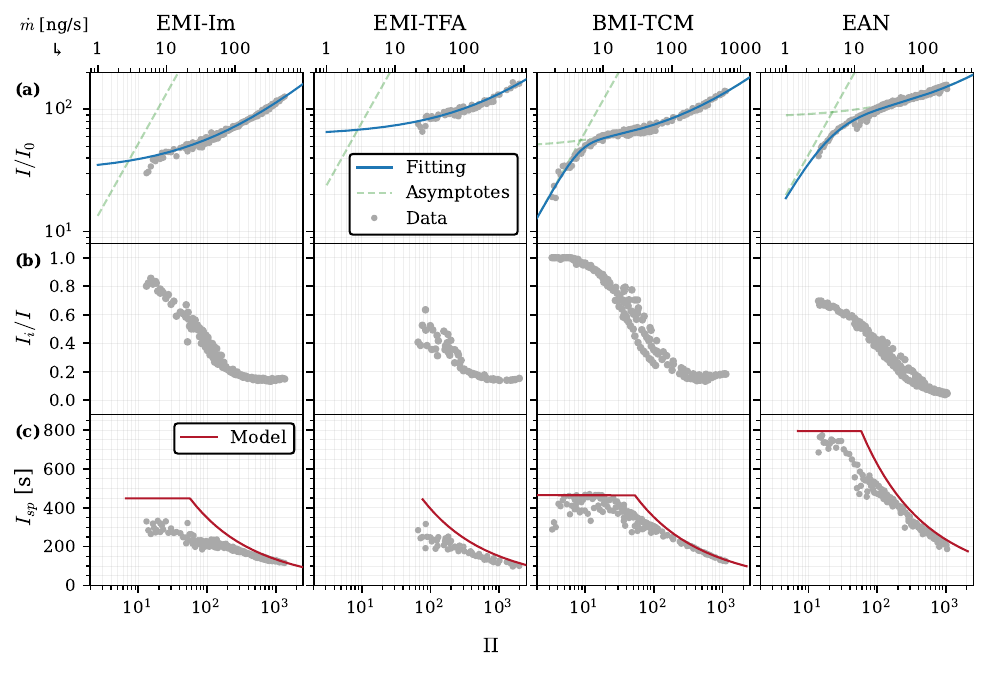}
    \caption{(a) Dimensionless current versus flow rate. The dashed asymptotes correspond to the terms $I/I_0=\Lambda \Pi$ and $I/I_0=\nu + \psi \Pi^{1/2}$. Dissociation fractions used: $\alpha=0.095$ for EAN, $\alpha=0.13$ for BMI-TCM, $\alpha=0.15$ for both EMI-Im and EMI-TFA. (b) Dimensionless ion current versus flow rate. (c) Specific impulse experimental data (corrected for bias) and model. The dissociation fractions used for the latter are the same as row (a).}
    \label{fig:curr_flow}
\end{figure*}

\subsection{Current scaling for highly conductive liquids. The dissociation limit}

The four highly conducting liquids tested follow the current scaling of Eq.~\eqref{eq:curr_scaling} across most flow rates with an added offset $I/I_0 \cong \nu + \psi \Pi^{1/2}$. The offset is well fitted by $\nu \sim Re^{-u}$ with $u\approx5/6$. This scaling suggests that the offset becomes significant only at very low Reynolds numbers (it is only larger than the classical square-root term at the minimum flow rate for $Re\lesssim\psi^{-3}\approx0.06$). It is also known to depend on the dielectric constant, though to a lesser extent \cite{Gamero-Castano2010a}, and its exact functional form requires further investigation.

\begin{figure*}[t!]\centering
    \includegraphics[width=\linewidth]{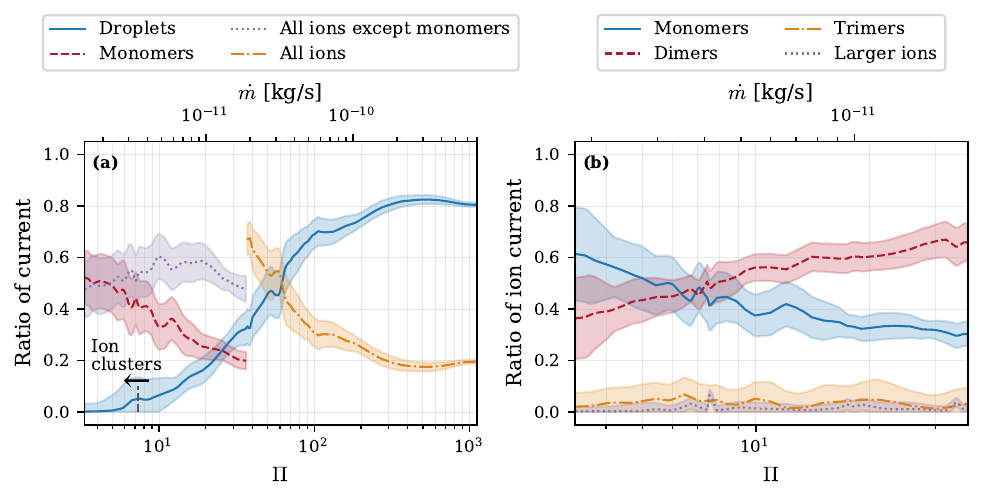}
    \caption{(a) Current fractions of beam components versus dimensionless flow rate $\Pi$. (b) Ion solvation state composition normalized to the total ion current versus $\Pi$.}
    \label{fig:ratio_currents}
\end{figure*}

At low flow rates, BMI-TCM and EAN deviate from this scaling and asymptotically approach a regime where current becomes proportional to mass flow rate: $I \cong \dot{m}/\xi_d$, with $\xi_d$ representing a minimum mass/charge. Overall, the dimensionless current law fits well to a smooth-minimum function of these two scalings:

\begin{equation}
    \frac{I}{I_0} \cong \left[\left(\Lambda \Pi\right)^{-z} + \left(\nu + \psi \Pi^{1/2}\right)^{-z}\right]^{-1/z} \label{eq:current_fitting}
\end{equation}

where $z$ controls the transition sharpness between the two regimes, $\Lambda=\xi_c/\xi_d$, and $\xi_c = \sqrt{\rho\varepsilon_0}/K$ is a mass/charge scale.

The fixed mass/charge regime is consistent with a supply-limited operational mode we term the dissociation limit. In the bulk ionic liquid upstream of the emission zone, quasi-neutrality holds (cation and anion number densities are equal, $n^+ \cong n^- \cong n$), and charge transport is dominated by ohmic conduction. The ions exist in dynamic equilibrium between free states contributing to conductivity and bound neutral pairs \cite{Feng2019}. The bulk dissociation fraction $\alpha$ is defined as the ratio of the number density of free anions or cations $n_f$ to the total number density of anions or cations $n$ at equilibrium:

\begin{equation}
    \alpha = \frac{n_f}{n}
\end{equation}

It is important to draw a subtle but crucial distinction between this free ion fraction and the frequently used macroscopic metric of ``ionicity.'' Ionicity is conventionally defined as the ratio of experimentally measured molar conductivity to the theoretical conductivity calculated from self-diffusion coefficients using the standard Nernst-Einstein equation \cite{MacFarlane2009, Nordness2020}. However, the straightforward application of the Nernst-Einstein equation to pure ionic liquids systematically overestimates conductivity \cite{Feng2019}. This discrepancy arises because the experimental diffusion coefficients account for the motion of all ion states, including those trapped in non-conducting neutral clusters. Instead, the free ion fraction represents the fraction of dissociated ions capable of moving independently and participating in DC conductivity at any given moment. When a modified Nernst-Einstein equation is formulated using exclusively the diffusion coefficients and concentrations of these free ions, the theoretical predictions achieve excellent agreement with experimental conductivity data~\cite{Feng2019}. For this reason, we employ the term free ion fraction to denote the physical fraction of available charge carriers, distinct from the macroscopic ionicity metric.

The relaxation time of the exchange between bound and free states is about 50~ps at 300~K and decreases with increasing temperature. Over the relevant temperature range, it is comparable to the electric relaxation time of the fluid at the current crossover point. For instance, for BMI-TCM at $\Pi = 18$, the liquid temperature and conductivity at the crossover are 415~K and 5.4~S/m, leading to an electric relaxation time of $t_k = 21$~ps; at this temperature, the relaxation time of the bound state is also about 20~ps. Because the timescales are comparable, dissociation cannot replenish free ions as fast as they are depleted by the electric field near the cone-jet, making the free-ion supply rate-limiting in the transition region.

In the limiting scenario where the exchange process between free and bound ions is frozen relative to the electrical relaxation time, the maximum flux of free ions available for extraction is $\dot{n}_f=\alpha \dot{m}N_A/M_p$. In this case, the entire flux of available free ions of the emission polarity (e.g., free cations) supplied by the bulk flow is fed to the cone-jet. The counter-ions (free anions) drift upstream and react electrochemically at the electrode interface to satisfy global charge conservation when operating the electrospray continuously in one polarity. In the process, the counter-ions produce neutral reaction products which may accumulate on the emitter electrode, evaporate, or be ejected with the spray. The exact reaction pathways and products are not well understood but are known to depend on the electrode material and liquid \cite{C.Kroon2006}. In any case, the emitted current is thus limited by the supply of free ions, $|I| \approx e \dot{n}_f$.

If the emission polarity is alternated periodically, faradaic currents may be prevented and instead the counter-ions charge a double layer at the emitter electrode. Upon polarity reversal, these accumulated ions are emitted alongside the incoming convective flux. Consequently, to maintain mass conservation over the full switching cycle, the instantaneous current magnitude during the active pulse must double relative to the continuous supply rate (if the emitted current is equal in both polarities). The minimum, or dissociation, mass/charge is then:

\begin{equation}
    \xi_d = \frac{\dot{m}}{|I|} = \frac{M_p}{\alpha \beta e N_A} \label{eq:xi_s}
\end{equation}

where $\beta=1$ corresponds to fixed-polarity operation and $\beta=2$ to switching-polarity operation. Above this limit, the beam's average mass/charge increases because a fraction of the emitted mass consists of bound neutral pairs that cannot contribute to the current. Note that operating at the dissociation limit does not preclude droplet emission; free ions may be carried within droplets or evaporate via field emission.

Fitting the experimental current data with Eqs.~\eqref{eq:current_fitting} and \eqref{eq:xi_s} yields dissociation fractions and mixing parameters \{$\alpha\approx0.13$, $z\approx3.5$\} for BMI-TCM and $\{\alpha\approx0.095$, $z\approx1.55$\} for EAN, and the resulting fits are compared with experimental data in Fig. \ref{fig:curr_flow}(a). A study using molecular dynamics simulations of highly conductive ionic liquids (including EMI-Im) determined the free ion fraction using multiple analytical criteria \cite{Feng2019}. They found it to be $\alpha\cong0.15$ at 295~K for all the liquids they modeled, which is in good agreement with our experimental fits. The authors argue that the dissociation fraction is expected to be similar for all pure ionic liquids.

\begin{figure*}[t!]\centering
    \includegraphics[width=\linewidth]{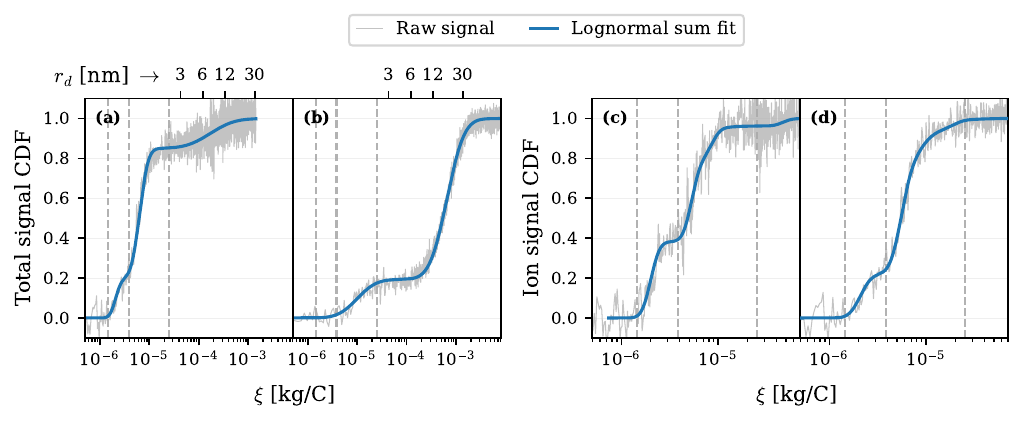}
    \caption{Panels (a1) and (a2): Total signal CDF as a function of mass/charge, raw signal (fine black line) and fitted lognormal sum (thick blue line). (a1) and (a2) correspond to $\Pi = 18.1$ and $\Pi = 376$, respectively. Panels (b1) and (b2): Ion signal CDF. (b1) and (b2) correspond to $\Pi = 6.2$ and $\Pi = 18.1$, respectively. Vertical dashed lines correspond to ions of solvation state $n=[0,1,10]$ from left to right. }
    \label{fig:fitted_tofs}
\end{figure*}

\subsection{Total current and ion current trends}

The total current $I$ and ion current $I_i$ as a function of flow rate are shown in Fig. \ref{fig:curr_flow}(b). The electrospray beam composition exhibits three distinct regimes as a function of dimensionless flow rate $\Pi$, with similar trends across all liquids. At high flow rates ($\Pi > 450$), the spray operates in a droplet-dominated regime where droplets carry approximately 80\% of the beam current while ions maintain a nearly constant fraction of 20\%. This also holds for EAN, although in Fig. \ref{fig:curr_flow}(b) its ion current appears lower than that of other liquids at high flow rates. This is because we define ion current $I_i$ as the portion of the current with a time-of-flight shorter than an arbitrary cutoff, here set at solvation state $n=10$. For the other liquids, this threshold cleanly separates the ions from droplets\cite{Caballero-Perez}, but due to EAN's small constituent ions, the evaporated ions have higher solvation states and wider distributions. By fitting a sum of lognormal distributions to discern ions from droplets, we verify that for EAN, too, approximately 80\% of the beam corresponds to a droplet distribution, while the remaining 20\% is due to ions and likely smaller droplets ejected from Rayleigh fission. The ion and small-droplet current fraction observed at high flow rates for all liquids is consistent with the approximately 20\% ejected during Rayleigh fission and evaporation at jet breakup, as detailed in Section \ref{sec:jet_breakup}. In this flow rate range, the current law closely follows the scaling $I/I_0=\nu + \psi \Pi^{1/2}$.

In the intermediate flow rate range ($50 < \Pi < 450$), a regime emerges with the ion fraction progressively increasing as the flow rate decreases. At about $\Pi = 80$, ions and droplets contribute equal fractions of the current. The current scaling remains the same for all liquids except EAN, which starts transitioning into the dissociation-limit scaling $I/I_0=\Lambda \Pi$ at $\Pi\approx100$. This transition occurs at lower flow rates for BMI-TCM.

Finally, at low flow rates ($\Pi < 50$), the beam transitions toward ion-dominated emission. In the particular case of BMI-TCM, this regime culminates in pure ion emission with no droplets present between $\Pi = 4.2$ and $\Pi=7.3$. At about $\Pi = 20$, it transitions into the dissociation-limit current trend.

\begin{figure*}[t!]\centering
    \includegraphics[width=\linewidth]{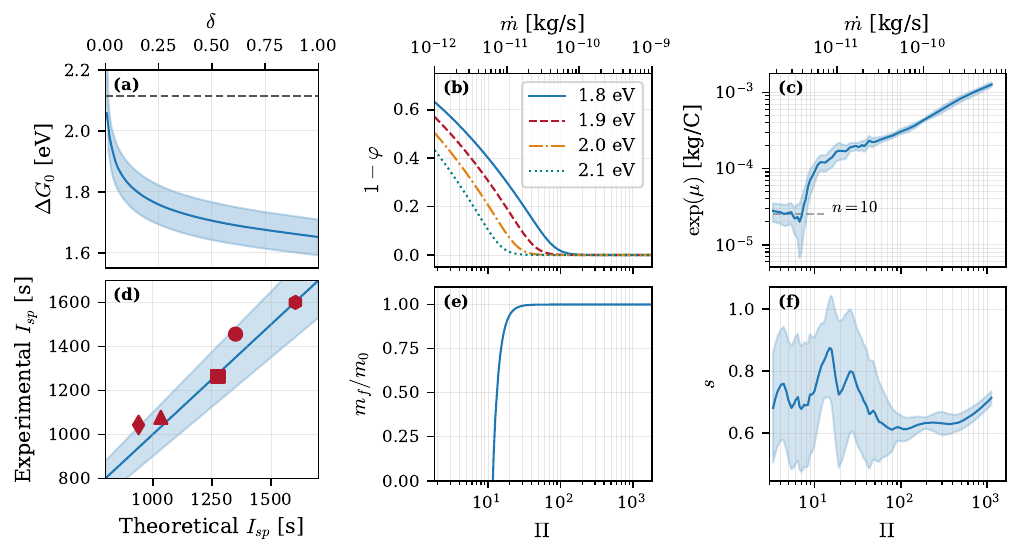}
    \caption{(a) $\Delta G_0$ estimate versus the fraction of ion current evaporated during flight, $\eta$. (b) Fraction of current evaporated throughout the jet for various values of $\Delta G_0$. (c) Median mass/charge $\exp(\mu)$ of the droplet lognormal component versus $\Pi$; the dashed line marks the $n=10$ ion cluster threshold. (d) Maximum theoretical specific impulse from Eq. \eqref{eq:ispmax} versus experimentally determined values. Shaded region indicates agreement within $\pm10\%$. Markers -- Circle\cite{Natisin2021}: EMI-BF$_4$ ($M_p=0.1980$ kg/mol), porous emitter, 2017~V; Square\cite{Huang2021}: EMI-BF$_4$, porous emitter, 1800~V; Triangle\cite{DeSaavedra2025}: EMI-Im, externally wetted emitter, 2000~V; Hexagon\cite{Demmons}: EMI-Im, unknown emitter type, 4800~V ($I_{\text{sp}}=1600$~s is the lowest uncertainty measurement reported at its maximum voltage); Diamond\cite{Galobardes-Esteban2026}: EAN, capillary emitter, 2099~V. (e) Final-to-initial critical droplet mass ratio after neutral evaporation over a transit time $t^*$. (f) Shape parameter $s$ of the droplet lognormal distribution versus $\Pi$.}
    \label{fig:dg0}
\end{figure*}

\subsection{Self-similarity of the droplet distribution}

Henceforth, we focus on BMI-TCM data as it exhibits a continuous evolution through all three regimes, including the pure ionic regime. The general trends presented here apply to the other liquids, except that only BMI-TCM reached the pure ionic regime. We obtained 832 TOF signals for BMI-TCM spanning mass flow rates from $\Pi=4.226$ to $\Pi=1425$ (1.81~ng/s to 611~ng/s). Unlike our previous study \cite{Caballero-Perez}, in the present study TOF signals at identical flow rates are treated individually rather than averaged and are not numerically filtered. Each TOF signal was mapped to a mass/charge cdf $F(\xi)$ and fitted to a sum of $N$ lognormal cdfs:

\begin{align}
    F(\xi) &\approx  \sum_{j=1}^N k_j Y_{j}(\xi;\mu_j, s^2_j) \label{eq:lognormal_mix} \\
    Y(\xi;\mu,s^2) &= \frac{1}{2}\left[1+\text{erf}\left(\frac{\ln \xi - \mu}{s\sqrt{2}}\right)\right]
\end{align}

The $3N-1$ fitting parameters comprise current fractions $k_j$ (summing to unity) and lognormal parameters $\mu_j$, $s_j$. For lognormal distributions, the median equals $e^\mu$, while the mean $m$ and coefficient of variation CV are:

\begin{equation}
    m=\exp\left(\mu+\frac{s^2}{2}\right),\qquad \text{CV} = \sqrt{\exp(s^2)-1} \label{eq:lognormal_mean}
\end{equation}

We obtain the optimal fitting parameters with maximum likelihood estimation in log-space. An analysis in Appendix \ref{appendix:mapping} shows that neglecting beam broadening and energy losses in the TOF to mass/charge mapping of Eq.~\eqref{eq:map} overestimates the actual median of the droplet mass/charge distribution by a factor of approximately 1.3--1.5. This factor increases with flow rate, as the beam becomes broader and the energy losses are higher (see Eq. \eqref{eq:delta_v}), increasing the error of the ideal mapping. Interestingly, its effects on $s$ are small, so the $s$ parameters we report are closely representative of the actual mass/charge distributions.

Figure~\ref{fig:ratio_currents}(a) shows the evolution of current fractions across flow rates, revealing the continuous transition between regimes. For $\Pi > 37$, two lognormal components are fitted: one for droplets and one for the combined ion population. Different ion solvation states cannot be distinguished at these flow rates due to signal overlap. For $\Pi < 37$, the monomer part of the signal is distinctly separated from other ions by a visible step between the expected TOFs of monomers and dimers, enabling a fit with three lognormal components. Figure \ref{fig:fitted_tofs}(a1) shows a TOF signal and the sum-of-lognormals fit with its visible monomer step; in contrast, Fig. \ref{fig:fitted_tofs}(a2) shows a signal obtained at a higher flow rate without the monomer step. This improved resolution allows detailed analysis of ion composition, which we explore in Section \ref{sec:ion_composition}.

In Fig. \ref{fig:dg0}(c), we show the median mass/charge of the droplet lognormal component. As expected, it decreases with decreasing flow rate, and at approximately $\Pi =7.3$ the median corresponds to an $n=10$ solvated ion cluster, arbitrarily marking the transition from droplets to large clusters. At that point, we can consider that the electrospray is emitting in the pure ionic regime.

A remarkable finding is the self-similar nature of the breakup process. Despite the median spanning nearly two orders of magnitude, the distribution's $s$ parameter, shown in Fig.~\ref{fig:dg0}(f), remains approximately constant in the range 0.7--1.1. This indicates highly skewed lognormal distributions with a constant coefficient of variation. This self-similarity, observed across all tested liquids, likely results from breakup consistently occurring in similar limiting conditions throughout the tested flow range: jet breakups characterized by high viscosity, high electrification, and equipotential evolution.

\subsection{Ion solvation state composition}\label{sec:ion_composition}

\begin{table}[h]
\caption{\label{tab:cluster_comparison}Comparison of ion cluster distributions reported in different studies.}
\begin{ruledtabular}
\begin{tabular}{lccccc}
 \begin{tabular}{@{}l@{}}Study \\ Liquid\end{tabular} &
 \begin{tabular}{@{}c@{}}This study \\ BMI-TCM\end{tabular} &
 \begin{tabular}{@{}c@{}}Corrado\cite{Corrado2024} \\ EMI-BF4\end{tabular} &
 \begin{tabular}{@{}c@{}}Krejci\cite{Krejci2017} \\ EMI-BF4\end{tabular} &
 \begin{tabular}{@{}c@{}}Natisin\cite{Natisin2021} \\ EMI-BF4\end{tabular} &
 \begin{tabular}{@{}c@{}}De Saavedra\cite{DeSaavedra2025} \\ EMI-Im\end{tabular} \\
\hline
 Ions & 18--100\% & 79--97\% & 94--97\% & 100\% & 97--99\% \\
 \quad $n=0$      & 30--62\% & 21--55\% & 29--39\% & 56--57\% & 35--58\% \\
 \quad $n=1$        & 37--67\% & 41--44\% & 43--54\% & 37\% & 36--45\% \\
 \quad $n=2$       & 6--16\%  & 3--10\%   & 7--11\% & 4--5\%  & 5--13\% \\
 \quad $2<n<10$  & 0--3\%      & 1--5\%      & 3--5\%  & 1--3\%  & 1--4\% \\
\hline
 Droplets  & 0--82\% & 3--11\% & 3--6\% & 0\% & 1--3\% \\
\end{tabular}
\end{ruledtabular}
\end{table}

For $\Pi\lesssim 37$, where monomers can be unambiguously distinguished from the TOF signals, we initially fit three components: monomers, higher solvation state ions (combined), and residual droplets or large clusters. Identifying the monomer TOF signature provides a reference point for determining the TOF signatures of higher solvation states. Since all ion species originate from the same emission zones, they share common retarding potential and angular distributions. Under this assumption, the TOF $\tau_n$ of ions with solvation state $n$ and mass/charge $\xi_n$ relates to the monomer TOF $\tau_0$ by $\tau_n =\tau_0 \sqrt{\xi_n/\xi_0}$. Their cdfs follow:

\begin{equation}
F_{\tau_n}(\tau)  = F_{\tau_0}\left(\tau \sqrt{\frac{\xi_0}{\xi_n}}\right) \label{eq:cdf_ions}
\end{equation}

We then map the TOF to mass/charge using Eq.~\eqref{eq:map} to obtain the mass/charge cdf for each ion solvation state, and fit a sum of lognormal cdfs for all solvation states to the ion portion of the signal (truncated at the mass/charge of $n=20$ clusters).

Figure \ref{fig:ratio_currents}(b) shows the current fractions as functions of flow rate, normalized to the ion current fraction. Ions with $n\geq3$ contribute minimally; their combined current fraction is shown. Nearly all ion current is carried by monomers and dimers, with less than 10\% trimers and trace amounts of higher solvation states. The predominance of monomers and dimers is consistent with the Born model (see Fig. \ref{fig:j_psi}(c)) and with previous studies that directly determined ion composition by sampling a reduced solid angle of the beam (Table \ref{tab:cluster_comparison}).

The monomer-to-dimer ratio varies significantly with mass flow rate: at approximately $\Pi = 3.7$, monomers comprise 60\% of ion current while dimers carry the remaining 40\%; at $\Pi = 37$, monomers carry 30\% of the current while dimers carry 70\%. In contrast, no discernible trend was observed for larger solvation states. Figs.~\ref{fig:fitted_tofs}(b1) and (b2) show fits at two different flow rates illustrating these different monomer-dimer proportions. The increasing monomer fraction at lower flow rates is counterintuitive at first: lower flow rates produce higher liquid temperatures, which according to the Born model (see Fig. \ref{fig:j_psi}(c)) should favor larger solvation states, not smaller ones. This trend can be explained as follows. At the cone-jet neck, where a local maximum of the electric field occurs, the liquid has undergone negligible self-heating; most of it occurs downstream of that point \cite{Magnani2024}. As the flow rate decreases, the electric field at the neck increases in accordance with Eq.~\eqref{eq:e_conejet}, causing an increasing proportion of the ion current to evaporate from this cold upstream region before reaching the hotter parts of the jet and droplets. As a result, the ion distribution shifts toward smaller solvation states characteristic of lower temperatures.

\subsection{Estimation of the ion solvation energy}\label{sec:dg}

In this section, we estimate the ion solvation energy $\Delta G_0$ by analyzing the ion current of BMI-TCM at $\Pi=466$ ($\dot{m}=200$~ng/s). At this flow rate, the measured ion current fraction from the TOF signal is $k_i=18.7\%$ while the droplet radii are distributed around $r^*\approx14$~nm and charged about 1\% over their Rayleigh limit. As mentioned earlier, droplets charged 4\% above the Rayleigh limit in the radius range 9--23~nm shed 18--24$\%$ of their charge via Rayleigh ion evaporation immediately following breakup \cite{Misra2023}, which is very close to the total fraction of current carried by ions we observe. Additionally, studies measuring the ion emission point with EMI-Im at similar flow rates do not detect the faint signal of ions that would correspond to evaporation from droplets in flight \cite{Gamero-Castano2021}. This suggests that almost all the ion current we observe at this flow rate originates during the jet breakup.

However, because the energy barrier is finite, a small fraction $\delta$ of the total ion current detected must still originate from droplets during flight. Consequently, a portion $k_i(1-\delta)$ of the current is attributed to that shed during the breakup, while the remaining $k_i\delta$ evaporates during flight. We then treat $\delta$ as a free parameter to determine $\Delta G_0$, noting that its value must be small.

We compute the charge lost to ion evaporation from a family of droplets in flight by integrating the Iribarne-Thomson expression \eqref{eq:iribarne}. We first rewrite it as:

\begin{equation}
    \exp\left(-\frac{G}{k_B T}\right) \frac{dq}{q} = - \frac{k_B T}{h} \exp\left(-\frac{\Delta G_0}{k_B T}\right) dt
    \label{eq:dg_droplets}
\end{equation}

and note from Eq.~\eqref{eq:delta_gep} that the field-induced reduction in the barrier scales approximately as $G\propto q^{1/2}$. Since evaporated ions remove a negligible fraction of the droplet mass, the radius change is ignored and the reduction in the energy barrier after an evaporation time $t_e$ is related to the initial one by $G(t_e)=G|_i\,\varphi^{1/2}$, where $\varphi=q(t_e)/q|_i$ is the ratio of the final to the initial charge during the evaporation event. With this change of variables the integral of Eq.~\eqref{eq:dg_droplets} can be expressed using the exponential integral $\operatorname{Ei}(x)=\int_{-\infty}^x u^{-1}e^u\,du$:

\begin{equation}
    \operatorname{Ei}\!\left(-\frac{G|_i}{k_B T}\right) - \operatorname{Ei}\!\left(-\frac{G|_i\, \varphi^{1/2}}{k_B T}\right) = \frac{k_B T\, t_e}{2 h} \exp\!\left(-\frac{\Delta G_0}{k_B T}\right).
    \label{eq:iribarne_integrated}
\end{equation}

Given the initial energy barrier reduction, temperature, and the fraction of the charge $1-\varphi$ evaporated as ions in time $t_e$, Eq.~\eqref{eq:iribarne_integrated} can be solved for $\Delta G_0$. To display the leading dependence on the variables, for small charge changes ($(1-\varphi) \ll 1$) one may expand $\operatorname{Ei}(x\varphi^{1/2}) \approx \operatorname{Ei}(x) + e^x(\varphi^{1/2}-1)$ and obtain the first-order approximation

\begin{equation}
    \Delta G_0 = G|_i + k_B T \ln\!\left( \frac{k_B T\, t_e}{2 h (1 - \varphi^{1/2})} \right).
    \label{eq:iribarne_int_simple}
\end{equation}

Equation~\eqref{eq:iribarne_int_simple} shows that the determination of $\Delta G_0$ is most sensitive to the estimate of $G|_i$, which scales roughly with the square of the droplet electric field prior to evaporation. By contrast, the result is relatively insensitive to $t_e$ and to small changes in $\varphi$. We obtain the distribution of $G|_i$ by first computing the radius and charge of the critical droplets with Eqs.~\eqref{eq:rstar} and \eqref{eq:estar} without considering the charge lost to ion evaporation at the breakup. After a fraction $k_i(1-\delta)$ of the charge is lost, the critical droplet electric field becomes $E_2^* = E^* \bigl(1-k_i(1-\delta)\bigr)$, while the droplet radius $r^*$ remains approximately unchanged. This updated value of the critical droplet electric field is used to obtain the distribution of a droplet family of a given mass/charge with Eq. \eqref{eq:r_E_distr}. The droplet radius distribution is also obtained with Eq.~\eqref{eq:r_E_distr}. Finally, $G|_i$ can be obtained for each droplet family with Eq. \eqref{eq:G}. For reference, the reduction in the barrier for the critical droplet at this flow rate is $G|_i(\xi_j)=1.13\ \mathrm{eV}$.

Note that we are implicitly assuming that all droplet families lose the same fraction $k_i(1-\delta)$ of their charge at breakup. This is a reasonable assumption, as most droplets in this case are within a range of 9-24~nm (see Fig. \ref{fig:fitted_tofs}(a2)), and within that range the portion of the charge lost to Rayleigh ion evaporation remains practically constant \cite{Misra2023}. Smaller droplets, owing to their larger electric fields and $G|_i$, will evaporate a larger fraction of their charge during flight. The remaining charge fraction of a droplet family after flight $\varphi(\xi)$ is obtained from solving Eq.~\eqref{eq:iribarne_integrated} for a given $\Delta G_0$. Extending this calculation over the full droplet distribution yields the consistency condition that the total fraction of current evaporated in flight equals $k_i\delta$:

\begin{equation}
    1-\int_0^\infty \varphi(\xi)\, f_d(\xi)\, d\xi = k_i\delta,
    \label{eq:dg0_fitting}
\end{equation}

For the mass/charge pdf, we use a lognormal distribution with mean mass/charge of $\dot{m}/\left[I(1-k_i\delta)\right]$ and the experimentally obtained $s_d=0.6$ (the mean and $s_d$ parameter together determine $\mu_d$). Solving Eq.~\eqref{eq:dg0_fitting} for $\Delta G_0$ as a function of $\delta$ gives the curve plotted in Fig.~\ref{fig:dg0}(a). The figure also shows the Born energy barrier estimate, the sensitivity bands corresponding to $\pm 10\%$ errors in the droplet electric field, the result with $\zeta(\kappa)=1$ (neglecting curvature effects), the estimate using only the critical droplet rather than the full distribution, and the solution for an evaporation time 3 times larger than $t_e$.

The unphysical scenario where $\delta\rightarrow0$, which implies that all ions originate from the breakup and none evaporate at flight, requires an infinitely high energy barrier. Because we cannot quantify the minimum value of $\delta$, we cannot obtain the upper threshold of $\Delta G_0$ with this method. In the other extreme where $\delta=1$, implying that all ions evaporate during flight, $\Delta G_0 = 1.64$~eV, which provides the lower bound for the ion solvation energy. A reasonable range for $\delta$ is likely about $\delta \lesssim5\%$ for the reasons that were mentioned earlier, yielding $\Delta G_0\gtrsim1.9$~eV. This is close to the analytical estimate of $\Delta G_0 = 2.1$~eV obtained with Eq.~\eqref{eq:dg_born}.

Notably, the relatively high value of the lower-bound $\Delta G_0 = 1.64$~eV and the likely values of $\Delta G_0\gtrsim 1.9$~eV make it difficult to reconcile with the notion of a jet-less ion-emitting cone when electrosprays operate in the pure ionic regime. For instance, a hemispherical cap that emits an ion current $I$ and whose electric and surface stresses are balanced yields $
    \Delta G_0
= \left[
(K \gamma^{2} e^{9})/(
     4\pi^{2}\varepsilon\,\varepsilon_0^{5} I)
\right]^{1/6}
$ \cite{Coffman2019}. The ion current $I=100$~nA we observe in the pure ionic regime with BMI-TCM would require a much lower value of the energy barrier under this hypothesis, $\Delta G_0=1.25$~eV. In a more detailed numerical study that also assumed that ions evaporate from a jet-less cone tip, the upper limit of $\Delta G_0$ for several liquids was obtained, and was found to be 1.48~eV and 1.64~eV for EMI-Im and EMI-BF$_4$, respectively \cite{Magnani2023}. All these values are considerably lower than our estimate. Additionally, the presence of a jet and droplets may help explain the large propellant losses that occur at low flow rates, which will be explored in the next section.

As the flow rate decreases, the electric field at the cone-jet increases (Eq. \eqref{eq:e_conejet}). Below a certain threshold flow rate, the electric field becomes sufficient to evaporate ions from the jet. The liquid residence time at the jet is approximately:

\begin{equation}
    t_j \sim \frac{l_j}{v_j} \sim \frac{\pi \rho l_j r_j^2}{\dot{m}}
\end{equation}

Using Eq. \eqref{eq:delta_gep} to obtain the reduction in the energy barrier at the jet with $t_j$ as the evaporation time, we can solve for $\varphi$ in Eq. \eqref{eq:iribarne_integrated}. In Fig. \ref{fig:dg0}(b), we show the fraction of the current evaporated at the jet $1-\varphi$ versus the flow rate for various values of $\Delta G_0$ and $\tilde{r}_j=0.28$. Although we cannot determine the exact flow rate at which ion evaporation from the jet becomes significant, for reasonable values of $\Delta G_0=1.9$--2.2~eV ion evaporation from the jet occurs well before the minimum flow rate.

\section{Electrospray thruster performance}\label{sec:propulsion}

In this section we discuss aspects of highly conductive electrosprays that are relevant to propulsion, such as the propulsive efficiency and specific impulse. A phenomenon affecting performance is the loss of propellant as neutral species. In our previous study, we observed that the ratio between charged particle mass flow (measured via time-of-flight) and total supplied mass flow (measured directly) decreases significantly at flow rates below $\Pi = 18$ for all tested liquids. This discrepancy between indirect and direct flow rate measurements, first reported by \cite{Natisin2021} and subsequently confirmed~\cite{DeSaavedra2025}, represents a primary inefficiency source at low flow rates that exceeds beam broadening or energy losses. It also renders indirect characterization methods unreliable for determining mass flow rate and specific impulse. Before examining the mechanisms underlying these losses, we define the relevant propulsion performance metrics.

We denote charged mass flow rate as $\dot{m}_{\text{ch}}$ to distinguish it from total supplied flow $\dot{m}$. Charged particles in the interval [$\xi$, $\xi+d\xi$] carry a current differential $dI_{\text{ch}}  =  I\,f(\xi)d\xi$. The charged particle mass flow rate is:

\begin{equation}
    \dot{m}_{\text{ch}} = \int \xi \, dI_{\text{ch}} = I \,\int_0^\infty  \xi f(\xi) \, d\xi = I \,\langle\xi\rangle
    \label{eq:mdotj}
\end{equation}

where $\langle\xi\rangle$ is the mean mass/charge. Only charged particles contribute to thrust (energetic neutral particles from droplet evaporation are neglected). Neglecting beam broadening and energy losses, particles with mass/charge $\xi$ achieve axial speed $c(\xi)=\sqrt{2V/\xi}$. Integration over all species yields the following expression for thrust:

\begin{equation}
T \;=\; \int_{0}^{\infty} c(\xi)\, d\dot m_{\rm ch}(\xi)
 \;=\; \int_{0}^{\infty} \sqrt{\frac{2V}{\xi}} \, I\,\xi\, f(\xi)\,d\xi
 \;=\; I\sqrt{2V}\;\langle\xi^{1/2}\rangle.
\label{eq:thrust_simplified}
\end{equation}

Thruster efficiency comprises two primary components: utilization efficiency $\eta_u$ (fraction of total mass flow ejected as charged particles) and polydispersive efficiency $\eta_p$ (penalty for broad mass/charge distributions):

\begin{equation}
    \eta = \frac{ T^2 }{2 \dot{m} P_{\text{in}}} = \underbrace{\frac{\dot{m}_{\text{ch}}}{\dot{m}}}_{\eta_u} \underbrace{\frac{\langle\xi^{1/2}\rangle^2}{\langle\xi\rangle}}_{\eta_p}
\end{equation}

where $P_{\text{in}} = IV$ is the input power. The specific impulse therefore becomes:

\begin{equation}
    I_{\mathrm{sp}} = \frac{T}{g_0 \dot{m}}=\frac{\eta_u}{g_0} \sqrt{\frac{2V \eta_p}{\langle\xi\rangle}} \label{eq:isp}
\end{equation}

Although not derived here for the sake of brevity, additional efficiency terms include energy efficiency $\eta_e\approx1-\Delta V/V$, angular efficiency $\eta_\theta$, and transmission efficiency $\eta_t$ (accounting for beam impingement on electrodes) \cite{Lozano2005a}. These terms modify specific impulse by a factor $\eta_t \sqrt{\eta_e \eta_\theta}$. Typical values in the ion-dominated regime are $\eta_\theta=0.7$--0.9, $\eta_e>0.9$, and $\eta_t>0.98$ \cite{Natisin2021, Ramos-Tomas2024, Krejci2017}. While energy losses can become significant at high flow rates in the droplet regime, they can be mitigated by increasing acceleration voltage. Angular efficiency can be also improved with a higher acceleration voltage \cite{Gamero-Castano2022}. Consequently, utilization and polydispersive losses typically dominate overall inefficiency, especially in the ion-dominated regime.

\subsection{Neutral losses}
We explore three potential mechanisms that could generate neutral species: electrochemical reactions, thermal degradation, and vapor-phase evaporation. Analysis in Appendix \ref{appendix:neutral} demonstrates that electrochemical reactions and thermal degradation cannot account for the observed losses, leaving evaporation as the dominant mechanism. The evaporation rate into vapor follows the Schrage equation~\cite{Liang2017}:

\begin{equation}
    \dot{m}_n = \frac{2\hat{\sigma}}{2 - \hat{\sigma}}\, p A \left( \frac{M_p}{2 \pi R T} \right)^{1/2}
    \label{eq:dmdtneutral}
\end{equation}

where $M_p$ is the molar mass of an anion--cation neutral pair (as ionic liquids evaporate as such pairs~\cite{Horike2018}), $A$ is the evaporating area, and $\hat{\sigma}$ is the accommodation coefficient. The Schrage equation adds a correction factor to the Hertz-Knudsen equation that depends on $\hat{\sigma}$; we assume $\hat{\sigma} = 1$ based on previous studies~\cite{Shan2019}.

To identify the evaporation source, we evaluate potential locations: the Taylor cone, the jet, and the droplets. The Taylor cone contribution is negligible as significant self-heating occurs only downstream along the jet \cite{Magnani2024}, leaving the cone near room temperature where vapor pressure is too low for appreciable evaporation.

For the jet, with $A\sim 2\pi r_c l_j$ and Eq.~\eqref{eq:lj}, the ratio of neutral evaporation to total mass flow is:

\begin{equation}
    \frac{\dot{m}_{n}}{\dot{m}}\bigr|_\text{jet} \sim \tilde{l}_j \,p \left( \frac{\varepsilon_0}{\gamma \rho^2 K} \right)^{1/3} \left( \frac{8 \pi M_p}{R T} \right)^{1/2}.
    \label{eq:evap-jet}
\end{equation}

Even with conservative estimates ($\tilde{l}_j = 10^3$ and $p \approx \SI{10}{\pascal}$ at \SI{550}{\kelvin} for EMI-Im at minimum flow rate), this ratio remains negligible ($\dot{m}_{n}/\dot{m} < 10^{-3}$).

Droplet evaporation, however, becomes the dominant neutral loss mechanism below approximately $\Pi = 18$, coinciding precisely with the onset of observed mass flow discrepancies. Three compounding factors intensify droplet evaporation at low flow rates. First, decreasing flow rate intensifies cone-jet self-heating, elevating droplet temperatures and exponentially increasing vapor pressure. Second, the smaller droplets produced at lower flow rates experience enhanced vapor pressure due to curvature effects. Third, the increased surface-to-volume ratio of smaller droplets leads to higher relative mass loss rates.

Figure~\ref{fig:dg0}(e) quantifies this effect, showing the final-to-initial mass ratio of EMI-Im critical droplets after evaporation during acceleration region transit. The calculation incorporates the current fitting~\eqref{eq:current_fitting}, evaporation time~\eqref{eq:t_star}, temperature profile~\eqref{eq:tempincrease}, and critical droplet radius~\eqref{eq:rstar}. Similar results are obtained for other ionic liquids.

\subsection{Specific impulse in the droplet-dominated regime}
The self-similar droplet mass/charge lognormal distribution has direct implications for thruster performance. To evaluate these, we decompose the mass/charge distribution into ion and droplet components:

\begin{equation}
    f(\xi) =  (1-k_d) f_i(\xi) + k_d f_d(\xi) \label{xi_distrib}
\end{equation}

where $k_d$ is the droplet current fraction. When the mean droplet mass/charge greatly exceeds that of the ions, the moments are dominated by the droplet component ($\langle\xi^{1/2}\rangle \approx k_d\langle\xi_d^{1/2}\rangle$, $\langle\xi\rangle \approx k_d\langle\xi_d\rangle$), and the polydispersive efficiency simplifies to:

\begin{equation}
    \eta_p =\frac{\langle\xi^{1/2}\rangle^2 }{\langle\xi\rangle } \approx \frac{\left(k_d\langle\xi_d^{1/2}\rangle\right)^2 }{k_d\langle\xi_d\rangle }=k_d \exp\left(-\frac{s_d^2}{4}\right)
    \label{eq:polydisp}
\end{equation}

Polydispersive efficiency thus depends only on the droplet current fraction and the distribution width parameter $s_d$, both of which remain approximately constant across the droplet-dominated regime. The factor $\exp(-s_d^2/4)$ represents the inherent efficiency penalty from the lognormal distribution width.

Combining this polydispersive efficiency with the current scaling and efficiency components yields the specific impulse scaling for the droplet-dominated regime. At high flow rates, neutral losses are negligible. Substituting the high-flow-rate current scaling from Eq.~\eqref{eq:curr_scaling} and the polydispersive efficiency expression into Eq.~\eqref{eq:isp}:

\begin{align}
    I_\text{sp}|_{\text{drop}}&=\underbrace{\frac{1}{g_0}\sqrt{\frac{2V}{\xi_c}}}_{I_\text{sp}\,\text{scale}} \, \underbrace{\sqrt{\frac{Re^{-u}}{\Pi} + \frac{\psi}{\sqrt{\Pi}}}}_{\text{cone-jet factor}}\, \underbrace{\sqrt{k_d \exp\left(-\frac{s_d^2}{4}\right)}}_{\text{sqrt polydispersive eff.}} \, \underbrace{\sqrt{1-\frac{\Delta V}{V}}}_{\text{sqrt energy eff.}} \label{eq:isp_droplets}
\end{align}

Here, the cone-jet factor follows from the current scaling of Eq.~\eqref{eq:curr_scaling} with the offset $\nu \sim Re^{-u}$ introduced in Section~\ref{results}, and the energy efficiency term $\sqrt{1-\Delta V/V}$ accounts for the irreversible voltage losses discussed in Section~\ref{sec:theoretical_background}. The cone-jet factor, of order unity, decreases with increasing flow rate due to the current scaling. The polydispersive efficiency term remains nearly constant throughout the droplet-dominated regime: with $k_d\approx0.8$ (consistent with 20\% ion current from breakup, see Section~\ref{sec:jet_breakup}) and $s_d\approx0.6$--0.8, we obtain $\eta_p\approx0.7$.


\subsection{Maximum specific impulse}\label{sec:maxisp}

The dissociation limit establishes a fundamental ceiling for electrospray thruster specific impulse. As mass flow rate decreases toward this limit, the emission transitions to predominantly or exclusively ions of varying solvation states, a pattern observed not only in BMI-TCM but across most electrospray thrusters operating at comparable low flow rates.

An ion-dominated emission coincides with the dissociation limit due to various mechanisms. First, at dissociation-limit flow rates, our analysis indicates substantial ion evaporation directly from the cone-jet itself. Second, the small droplets produced at these flow rates undergo copious neutral evaporation, reducing their size and increasing the electric fields at their surface. This creates a cascade effect as this drives further ion evaporation, ultimately depleting the droplet population of almost all charged species.

The Born model provides an estimate for the mean emitted mass/charge: $\langle\xi\rangle\approx \langle\xi_i\rangle\approx \tfrac{4}{3}\pi\rho a_c^3/e$, yielding values between monomer and dimer mass/charge for typical large-molecule ionic liquids. Critically, this evaporation-determined mass/charge is much lower than the supply-limited value $\xi_d$ at the dissociation limit. Consequently, while all free ions evaporate, the remaining bound ions cannot be emitted as charged species. We can speculate that these bound ions are either ejected as low-speed neutral droplets from the jet or accumulate at the emitter, neither contributing to thrust. The utilization efficiency therefore becomes:

\begin{equation}
    \eta_u =\frac{\dot{m}_\text{ch}}{\dot{m}} \approx \frac{\langle\xi_i\rangle}{\xi_d}= \frac{\rho\,\alpha\,\beta\,N_A\,e^2\left(1-1/\varepsilon\right)}
{48\pi\,\gamma\,\varepsilon_0\,M_p}
\label{eq:etau}
\end{equation}

Reference \citenum{Natisin2021} measured the individual efficiency components in an electrospray thruster and found $\eta_u=0.394$ for their EMI-BF$_4$ thruster which also operated in the pure ion regime. Using $\alpha=0.13$, Eq.~\eqref{eq:etau} yields $\eta_u=0.343$, close to their experimentally determined value.

In the ionic regime, the narrow solvation state distribution (predominantly monomers and dimers) yields high polydispersive efficiencies of $\eta_p\approx0.8$--0.9 \cite{Natisin2021, MacArthur2024}. As explained previously, energy, angular, and transmission efficiencies remain high compared to the utilization efficiency. Since the latter dominates the losses, we obtain an expression for maximum specific impulse by neglecting secondary efficiency terms:

\begin{equation}
    I_{\text{sp}}\big|_{\text{max}} \approx \frac{\alpha \beta N_A}{g_0 M_p}\,
\sqrt{\frac{\rho Ve^{3}\bigl(1-1/\varepsilon\bigr)}{24\pi\,\gamma\,\varepsilon_0}} \label{eq:ispmax}
\end{equation}

In addition to the simplifications that have been described, Eq. \eqref{eq:ispmax} also assumes that electrochemical losses are negligible in switching-polarity thrusters. These may be avoided by applying the voltage to the liquid through an electrode distal to the emitter and by switching the emission polarity prior to the onset of faradaic currents. Appendix~\ref{appendix:isp} shows that Eq.~\eqref{eq:ispmax} underestimates the specific impulse for mixed ion-droplet beams at the dissociation limit, but the error diminishes rapidly with increasing ion current fraction and remains modest for predominantly ionic beams.

To validate our model, we compiled five independent studies with direct specific impulse measurements (both thrust and mass flow rate measured directly). Figure~\ref{fig:dg0}(d) compares their highest specific impulse values with predictions from Eq.~\eqref{eq:ispmax}. We use dissociation fractions of $\alpha=0.15$ for EMI-Im (from MD simulations \cite{Feng2019}), $\alpha=0.13$ for EMI-BF$_4$ (using BMI-TCM's value), and $\alpha=0.095$ for EAN (obtained in this study). Despite varying ionic liquids, emitter architectures, and acceleration voltages, the model remarkably agrees within $\pm10\%$ with all experimental values. For comparison, a naive estimate that assumes monomer emission with perfect efficiency gives $I_\text{sp} = (1/g_0)\sqrt{4VN_Ae/M_p}$, yielding 4527~s for EMI-Im at 2000~V. In contrast, Eq.~\eqref{eq:ispmax} predicts 1034~s, matching the experimentally measured 1078~s for that liquid and voltage \cite{DeSaavedra2025} within 4\%.

The model reveals pathways to higher specific impulse. Since dissociation fraction increases with temperature, elevated liquid temperatures should slightly enhance performance. However, reaching the dissociation limit requires sufficiently low flow rates, which can be achieved mostly by the use of highly conducting liquids and emitter geometries that produce small Taylor cones. Propellants with smaller molecular weight also lead to higher specific impulses, as well as switching-polarity operation, which doubles the effective ion flux per unit propellant mass.

In our previous study we determined the specific impulse using indirect thrust measurements via TOF with directly measured mass flow. This partially indirect method led to a systematic overestimation bias $\mathcal{B}\approx1.5$. Figure~\ref{fig:curr_flow}(c) shows experimental specific impulse (corrected for bias) compared with the minimum of dissociation-limit (Eq.~\eqref{eq:ispmax}) and droplet-regime (Eq.~\eqref{eq:isp_droplets}) predictions. The model uses the same parameters across liquids ($V=1500$~V, $\psi=2.5$, $k_d=0.8$, $s_d=0.7$, $u=5/6$, except dissociation fraction; see caption), and captures the observed behavior across all flow rates. Maximum deviation occurs at $\Pi=50$--200, where mixed-regime operation reduces polydispersive efficiency below the assumed values, as expected from the simplified two-regime model.

\section{Conclusions}\label{sec:conclusion}

We have investigated the physics and beam composition of electrosprays operating in the high-conductivity limit ($K\gtrsim1$ S/m), characterizing the transition from droplet-dominated to ion-dominated emission for four ionic liquids: EMI-Im, EMI-TFA, BMI-TCM, and EAN. By combining TOF spectrometry with direct flow rate measurements, we have mapped the emission regimes of these sprays and identified fundamental properties and constraints governing their operation.

In the droplet-dominated regime (dimensionless flow rates $\Pi\gtrsim50$), the jet breakup process exhibits a self-similar lognormal distribution. Despite the median droplet mass/charge spanning nearly two orders of magnitude at varying flow rates, the variance of the distribution relative to its mean remains constant, yielding a nearly constant polydispersive efficiency of $\eta_p\approx0.7$ throughout this regime. At flow rates above $\Pi>450$, ions represented 20\% of the beam's current for all the liquids, consistent with ion evaporation during and shortly after jet breakup, including Rayleigh fission of the newly formed droplets. As the flow rate decreases below $\Pi<450$, the beam composition shifts continuously toward ion-dominated emission, with the average ion solvation state decreasing. Contrary to expectations based on Born's model and the increasing liquid self-heating, monomers become the dominant species at the lowest flow rates. This suggests that the intensifying electric field at the cone-jet neck shifts the primary ion evaporation zone to this cooler upstream region, away from the hotter jet and droplets.

By modeling the ion evaporation energy barrier and the electric fields at the droplets, we estimated the lower bound for the ion solvation energy of BMI-TCM to be $\Delta G_0\gtrsim$1.9~eV, a value consistent with analytical estimates derived from the Born model. This high energy barrier suggests that a jet-less ion-emitting Taylor cone may be energetically infeasible even when the beam is solely comprised of ions.

We have identified two physical mechanisms that impose fundamental limits on the performance of ionic liquid electrosprays toward their minimum stable flow rate. First, a significant discrepancy arises between the charged mass flow rate inferred from TOF spectrometry and the total mass flow rate supplied to the electrospray. We attribute this to substantial neutral mass losses driven by the evaporation of small droplets, which are susceptible to rapid mass loss due to their high surface-to-volume ratios and the elevated temperatures of the jet. This neutral evaporation represents a primary efficiency loss mechanism in the high-specific impulse regime.

Second, we propose the existence of a dissociation limit determined by the free ion fraction of the bulk liquid. As the flow rate decreases, the current scaling deviates from the classical square-root law and asymptotically approaches a constant mass/charge regime defined by the availability of dissociated ions. This limit implies that the average mass/charge of the beam cannot be arbitrarily reduced, as bound neutral pairs are either ejected within droplets, carried as solvation shells around free ions, or lost as neutrals. The solvation state of the evaporated ions and thus the beam mass/charge is determined by the ion evaporation dynamics. When only ions are emitted, the discrepancy between the evaporated mass/charge of the ions and the minimum mass/charge of the dissociation limit leads to large propellant losses and a low utilization efficiency, dominating all other loss mechanisms. Based on the dissociation limit and this predicted efficiency, we derived an analytical expression for the maximum attainable specific impulse of an electrospray thruster. This expression agrees within $\pm10\%$ with five independent experimental datasets spanning different ionic liquids and emitter architectures, providing a physical ceiling for the capabilities of electrospray propulsion technology. The model also identifies pathways to higher specific impulse, including propellants with lower molecular weight, higher conductivity liquids that enable lower minimum flow rates, and switching-polarity operation.

\begin{acknowledgments}
This work was funded by the Air Force Office of Scientific Research, award number FA9550-21-1-0200, the Defense Advanced Research Projects Agency, award number HR00112490392, and fellowships from the Balsells Foundation and the William and Ida Melucci Space Exploration \& Technology Fellowship. M. Caballero-P\'erez would like to sincerely thank Marco Magnani, Marc Galobardes-Esteban, and Juan Fern\'andez de la Mora for their insightful discussions.\end{acknowledgments}

\section*{Conflict of Interest}
The authors have no conflicts to disclose.

\section*{Author Contributions}
\textbf{Manel Caballero-P\'erez:} Conceptualization (lead); Data curation (lead); Formal analysis (lead); Investigation (lead); Methodology (equal); Software (lead); Validation (lead); Visualization (lead); Writing -- original draft (lead); Writing -- review \& editing (equal). \textbf{Manuel Gamero-Casta\~no:} Funding acquisition (lead); Methodology (equal); Resources (lead); Project administration (lead); Supervision (lead); Writing -- original draft (supporting); Writing -- review \& editing (equal); Conceptualization (supporting).

\section*{Data Availability}
The data that support the findings of this study are available from the corresponding author upon reasonable request.

\section*{Appendix}
\appendix
\section{Analysis of the TOF to mass/charge mapping}\label{appendix:mapping}

From the TOF distribution, we wish to obtain the distribution in mass/charge. The mass/charge $\xi$ is a function of the retarding potential $V_r$, the ray angle $\theta$, and the TOF $\tau$:

\begin{equation}
    \xi = \left(\frac{\tau\cos \theta}{L}\right)^2 2V_r \label{eq:xi_full_app}
\end{equation}

Because in this study we do not directly measure the angular or retarding potential distribution, we will map TOF to an uncorrected estimate of the mass/charge $\hat{\xi}$:

\begin{equation}
    \hat{\xi} = \left(\frac{\tau}{L}\right)^2 2V \label{eq:xi_hat_app}
\end{equation}

We will now discuss how the estimated distribution $\hat{\xi}$ that we report differs from the actual distribution $\xi$. In cone-jet electrosprays, the current density is approximately constant for low polar angles and decreases rapidly above a certain threshold polar angle $\theta_m$, hence we can approximate $\theta \sim \mathcal{U}[0,\theta_m]$. Meanwhile, the retarding potential distribution can be simplified to be the emitter voltage minus a constant voltage drop $\hat{V}_r = V-\Delta V$. Taking the logarithm of Eqs.~\eqref{eq:xi_full_app} and \eqref{eq:xi_hat_app}:

\begin{equation}
    \ln \xi = \underbrace{2 \ln \tau + c}_{\ln \hat{\xi}} + \underbrace{\ln\!\left(\frac{\hat{V}_r}{V}\right)}_{A} + \underbrace{2\ln (\cos\theta)}_{B} \label{eq:ln_xi_decomp}
\end{equation}

where $c=\ln\left(\frac{2V}{L^2}\right)$ is a constant. The term $A = \ln(1-\Delta V/V)$ accounts for the retarding potential bias and is approximately constant across the beam (all particles experience the same voltage drop). The term $B$ captures the angular broadening. If $\ln\tau \sim \mathcal{N}(\mu_\tau, s_\tau^2)$ is normally distributed, $\ln \hat{\xi}$ is also a normal with:

\begin{align}
    \mu_{\hat{\xi}} &= 2\mu_\tau +c \\
    s_{\hat{\xi}}^2 &= 4s_\tau^2
\end{align}

Since $A$ is approximately constant (zero variance) and $B$ has a non-normal distribution, $\ln \xi$ is approximately normal if the variance of $B$ is much smaller than that of $\ln \hat{\xi}$:

\begin{align}
    \ln \xi &\sim \mathcal{N}(\mu_\xi,s^2_\xi) \leftrightarrow \xi \sim \text{Lognormal}(\mu_\xi, s^2_\xi)\\
    \mu_\xi&=\mu_{\hat{\xi}}+A+\langle B \rangle \\
    s^2_\xi &= s^2_{\hat{\xi}}+\text{Var}[B]
\end{align}

We can use the series approximation:

\begin{equation}
    2 \ln \cos \theta \approx -\theta^2 - \frac{\theta^4}{6}-\mathcal{O}(\theta^6)
\end{equation}

to obtain an expression for the expectation and variance of $B$, assuming $\theta \sim \mathcal{U}[0,\theta_m]$:

\begin{align}
    \langle B \rangle &= \int_0^{\theta_m} 2 \ln \cos(\theta) \frac{1}{\theta_m} d\theta=  -\frac{\theta_m^2}{3}-\frac{\theta_m^4}{30}-... \\
    \text{Var}[B] &= \langle B^2 \rangle-\langle B \rangle^2= \frac{4}{45}\theta_m^4 + ...
\end{align}

From fitting the experimental data, we find that $s^2_{\hat{\xi}}\sim0.35$--$0.65$, which is over an order of magnitude larger than the variance of $B$ for typical values of $\theta_m$ (about 30$^\circ$--45$^\circ$). This implies that the beam broadening adds a relatively small amount of variance to the $\ln \xi$ distribution, and $\ln\hat{\xi}$ provides a good estimate for its variance. In other words, the lognormal parameter $s$ obtained from mapping TOF to mass/charge neglecting beam broadening has a small error.

Both correction terms, however, shift the median. The median of $\xi$ relates to that of $\hat{\xi}$ through:

\begin{equation}
    \text{med}(\hat{\xi}) = \text{med}(\xi) \cdot \frac{1}{1-\Delta V/V}\cdot\exp\!\left(\frac{\theta_m^2}{3}\right) \label{eq:median_bias}
\end{equation}

The first factor arises from the retarding potential bias: the mapping uses the emitter voltage $V$, which exceeds the actual retarding potential $\hat{V}_r$ by the voltage drop $\Delta V$ in the cone-jet. The second factor is the angular broadening correction. For $\theta_m = \pi/6$, the angular factor alone is approximately 1.1. For the combined overestimation, using $\Delta V/V$ from Eq.~\eqref{eq:delta_v}, the total factor is approximately 1.3--1.5, increasing with flow rate as both the beam half-angle and the voltage losses grow.

To sum up, the estimate of the mass/charge distribution obtained with Eq.~\eqref{eq:xi_hat_app} preserves the shape parameter $s$ of the actual lognormal distribution with small error, but overestimates the median mass/charge by a factor of approximately 1.3--1.5 due to the combined effects of the retarding potential bias and beam angular broadening.

\section{Mixed-regime specific impulse in the dissociation limit}\label{appendix:isp}

In the mixed droplet-ion regime, the polydispersive efficiency is less than unity, and Eq.~\eqref{eq:ispmax} underestimates the actual specific impulse because it neglects the thrust contribution from droplets. We show here that this underestimation diminishes rapidly with increasing ion current fraction and is small for predominantly ionic beams. Starting from the definition of specific impulse and substituting Eqs.~\eqref{eq:thrust_simplified} and \eqref{eq:xi_s}:

\begin{equation}
    I_{\mathrm{sp}} = \frac{T}{g_0 \dot{m}}=\frac{\sqrt{2V}}{g_0} \frac{\langle\xi^{1/2}\rangle}{\xi_d} \label{eq:isp_annex}
\end{equation}

Equation~\eqref{eq:ispmax} replaces $\langle\xi^{1/2}\rangle$ with $\langle\xi_i\rangle^{1/2}$, an approximation that is exact when the beam consists entirely of ions with a narrow mass/charge distribution. To quantify the error in the mixed regime, we model the mass/charge distribution as a sum of two delta functions representing ions and droplets:

\begin{equation}
    f(\xi) = k_i \,\delta(\xi-\xi_i) + (1-k_i) \,\delta(\xi-\xi_d)
\end{equation}

\noindent where $\xi_i$ and $\xi_d$ are the average mass/charge of the ions and droplets, respectively, and $k_i$ is the ion current fraction. Computing the half-moment of this distribution:

\begin{equation}
    \langle\xi^{1/2}\rangle = k_i\,\xi_i^{1/2} + (1-k_i)\,\xi_d^{1/2}
\end{equation}

The ratio of the actual specific impulse~\eqref{eq:isp_annex} to the approximation from Eq.~\eqref{eq:ispmax} is:

\begin{equation}
    \mathcal{R} = \frac{\langle\xi^{1/2}\rangle}{\langle\xi_i\rangle^{1/2}} = k_i + (1-k_i)\sqrt{\frac{\xi_d}{\xi_i}}
\end{equation}

Since $\xi_d > \xi_i$, $\mathcal{R} \geq 1$ and Eq.~\eqref{eq:ispmax} always underestimates the specific impulse. The physical origin of this underestimation is that droplets, despite having lower charge-to-mass ratios than ions, still contribute to thrust, an effect neglected by the purely ionic approximation. For predominantly ionic beams ($k_i$ close to unity), $\mathcal{R}$ can be expanded as:

\begin{equation}
    \mathcal{R} \approx 1 + (1-k_i)\left(\sqrt{\frac{\xi_d}{\xi_i}} - 1\right)
\end{equation}

\noindent which approaches unity linearly with the droplet current fraction. For most ionic liquids at the dissociation limit, the beam is dominated by ions and $\mathcal{R}$ is close to unity.

EAN represents a limiting case, as it retains a larger droplet fraction at the dissociation limit due to its lower conductivity. With $k_i=0.783$ and $\xi_d/\xi_i = 7.32$, the underestimation factor is $\mathcal{R}=1.37$. However, this overestimate of the actual specific impulse relative to Eq.~\eqref{eq:ispmax} is partially offset by the secondary efficiency terms (angular efficiency $\eta_\theta$ and energy efficiency $\eta_e$) not captured in the simplified expression, which reduce the measured specific impulse by a factor $\sqrt{\eta_e \eta_\theta} \approx 0.8$--$0.9$. The net result is that Eq.~\eqref{eq:ispmax} remains a reasonable estimate even for mixed ion-droplet beams, as confirmed by the agreement with experimental data in Fig.~\ref{fig:dg0}(d).

\section{Thermal degradation and electrochemical reaction losses}\label{appendix:neutral}

\subsection{Electrochemical reactions}

In electrosprays operating without polarity alternation, electrochemical reactions occur at the interface between the liquid and the emitter to sustain the emission of charged species~\cite{Mora2000}. While some thruster configurations attempt to mitigate these reactions by intermittently reversing the emission polarity~\cite{Lozano2004}, complete avoidance of reactions remains uncertain \cite{Castro2009}.

In this study, electrosprays operate exclusively at a constant positive polarity. Consequently, electrons must flow from the ionic liquid to the emitter and toward the high-voltage supply to compensate for the emitted positive charges. Ionic liquids oxidize and decompose when subjected to voltages beyond their electrochemical windows. The exact degradation mechanisms for each liquid are not fully understood \cite{C.Kroon2006} and vary depending on the ionic liquid and electrode material. One possibility is that the cation or anion undergoes a sequence of reactions, eventually breaking down into small volatile molecules such as CO$_2$, H$_2$O, and NO$_3$ \cite{Fujiwara2020, Pieczyńska2015}. Certain ionic liquids, such as EMI-BF$_4$, may produce non-volatile byproducts that accumulate on the emitter as a dielectric film, hindering conductivity \cite{Lozano2004} or causing capillary blockages \cite{Lozano2003}. Additionally, the emitter's conductive material may undergo oxidation and degradation. Significant etching of tungsten emitters has been observed when voltage is applied directly to them, whereas using a distal electrode can prevent this issue \cite{Brikner2013}. The choice of capillary material also influences the predominant oxidation process; for instance, stainless steel facilitates liquid oxidation, while copper capillaries are more prone to oxidation themselves, contributing metallic cations to the liquid \cite{VanBerkel2001}. Even in positive polarity, reduction can occur alongside oxidation reactions, as long as the net difference between the oxidation and reduction currents equals the emitted current \cite{VanBerkel2001}. It is energetically unlikely for the counter-ion to react together with a cluster of neutrals, breaking and rearranging bonds to form new neutral substances, and no references support such a mechanism. Therefore, we assume that one anion reacts for each elementary charge emitted. For an electrospray emitting in positive polarity, assuming no oxidation of the metal electrode and that only the anion undergoes oxidation, anions with a mass/charge $\xi_\mathrm{B}$ are consumed at a rate

\begin{equation}
    \dot{m}_{\mathrm{B}} = I \,\xi_\mathrm{B} .
    \label{eq:evap_curr}
\end{equation}

Given the absence of clogging or residue in the emitters for all tested liquids, it is reasonable to assume that all the decomposition products evaporate. Because the current scales to a first order as $I \sim \dot{m}^{1/2}$, at large flow rates the evaporated mass flow rate is negligible compared to the total mass flow rate. At low flow rates these losses are not insignificant, but they do not account for the substantial discrepancies observed between direct and indirect measurements of the flow rate, assuming the hypotheses are valid. For instance, for BMI-TCM operating in positive mode at $\dot{m} = 3 \times 10^{-12}$\,kg/s, only 5\% of the total mass flow rate would correspond to evaporation of reacted anions.

\subsection{Thermal decomposition}
Volatile neutral compounds resulting from thermal decomposition may evaporate from droplets in flight. If evaporation occurs in the field-free zone between the extractor and the collector, it does not affect the TOF signal because the arrival time of a charged particle depends solely on its entry speed into this region. However, if neutral particle evaporation takes place within the acceleration region, it alters the TOF signature, as droplets that lose mass accelerate to higher speeds. Consequently, the TOF measurement yields an apparent mass/charge lower than the initial value of the parent droplet, leading to an underestimation of the mass flow rate and thrust.

Assuming first-order decomposition kinetics, the fraction of a droplet's initial mass $m_0$ remaining after a residence time $t^*$ in the acceleration region is:

\begin{gather}
    \frac{m(t^*)}{m_0} = \exp(-\mathcal{K}_{\text{dec}}t^*) ,\\
    \mathcal{K}_{\text{dec}}= \mathcal{K}_0 \exp\left(-\frac{E_A}{RT}\right),
\end{gather}

where $\mathcal{K}_{\text{dec}}$ is the thermal decomposition rate constant. For ionic liquids, the activation energy $E_A$ and the prefactor $\mathcal{K}_0$ are typically on the order of 100$\,$kJ/mol and $10^{10}$--$10^{20}\,$s$^{-1}$, respectively. For instance, for EMI-Im, $E_A = 317\,$kJ/mol and $\mathcal{K}_0=1.2\times10^{20}\,$s$^{-1}$ \cite{Heym2015}. Even at the highest temperatures we expect the droplets to reach, the fractional mass loss is negligible: for EMI-Im at 700$\,$K and $t^*=1\,$\SI{}{\micro\second}, $\mathcal{K}_{\text{dec}}t^*\sim10^{-10}$, so the fraction of mass lost to decomposition is $1-\exp(-\mathcal{K}_{\text{dec}}t^*)\approx\mathcal{K}_{\text{dec}}t^*\sim10^{-10}$. Therefore, thermal decomposition does not significantly contribute to the propellant losses observed.

\bibliography{references}

\begin{thebibliography}{75}%
\makeatletter
\providecommand \@ifxundefined [1]{%
 \@ifx{#1\undefined}
}%
\providecommand \@ifnum [1]{%
 \ifnum #1\expandafter \@firstoftwo
 \else \expandafter \@secondoftwo
 \fi
}%
\providecommand \@ifx [1]{%
 \ifx #1\expandafter \@firstoftwo
 \else \expandafter \@secondoftwo
 \fi
}%
\providecommand \natexlab [1]{#1}%
\providecommand \enquote  [1]{``#1''}%
\providecommand \bibnamefont  [1]{#1}%
\providecommand \bibfnamefont [1]{#1}%
\providecommand \citenamefont [1]{#1}%
\providecommand \href@noop [0]{\@secondoftwo}%
\providecommand \href [0]{\begingroup \@sanitize@url \@href}%
\providecommand \@href[1]{\@@startlink{#1}\@@href}%
\providecommand \@@href[1]{\endgroup#1\@@endlink}%
\providecommand \@sanitize@url [0]{\catcode `\\12\catcode `\$12\catcode
  `\&12\catcode `\#12\catcode `\^12\catcode `\_12\catcode `\%12\relax}%
\providecommand \@@startlink[1]{}%
\providecommand \@@endlink[0]{}%
\providecommand \url  [0]{\begingroup\@sanitize@url \@url }%
\providecommand \@url [1]{\endgroup\@href {#1}{\urlprefix }}%
\providecommand \urlprefix  [0]{URL }%
\providecommand \Eprint [0]{\href }%
\providecommand \doibase [0]{http://dx.doi.org/}%
\providecommand \selectlanguage [0]{\@gobble}%
\providecommand \bibinfo  [0]{\@secondoftwo}%
\providecommand \bibfield  [0]{\@secondoftwo}%
\providecommand \translation [1]{[#1]}%
\providecommand \BibitemOpen [0]{}%
\providecommand \bibitemStop [0]{}%
\providecommand \bibitemNoStop [0]{.\EOS\space}%
\providecommand \EOS [0]{\spacefactor3000\relax}%
\providecommand \BibitemShut  [1]{\csname bibitem#1\endcsname}%
\let\auto@bib@innerbib\@empty
\bibitem [{\citenamefont {Fenn}\ \emph {et~al.}()\citenamefont {Fenn},
  \citenamefont {Mann}, \citenamefont {Meng}, \citenamefont {Wong},\ and\
  \citenamefont {Whitehouse}}]{Fenn1989}%
  \BibitemOpen
  \bibfield  {author} {\bibinfo {author} {\bibfnamefont {J.~B.}\ \bibnamefont
  {Fenn}}, \bibinfo {author} {\bibfnamefont {M.}~\bibnamefont {Mann}}, \bibinfo
  {author} {\bibfnamefont {C.~K.}\ \bibnamefont {Meng}}, \bibinfo {author}
  {\bibfnamefont {S.~F.}\ \bibnamefont {Wong}}, \ and\ \bibinfo {author}
  {\bibfnamefont {C.~M.}\ \bibnamefont {Whitehouse}},\ }\bibfield  {title}
  {\enquote {\bibinfo {title} {Electrospray {{Ionization}} for {{Mass
  Spectrometry}} of {{Large Biomolecules}}},}\ }\href {\doibase
  10.1126/science.2675315} {\ \textbf {\bibinfo {volume} {246}},\ \bibinfo
  {pages} {64--71}}\BibitemShut {NoStop}%
\bibitem [{\citenamefont {He}\ and\ \citenamefont {V.~Jokerst}()}]{He2020}%
  \BibitemOpen
  \bibfield  {author} {\bibinfo {author} {\bibfnamefont {T.}~\bibnamefont
  {He}}\ and\ \bibinfo {author} {\bibfnamefont {J.}~\bibnamefont
  {V.~Jokerst}},\ }\bibfield  {title} {\enquote {\bibinfo {title} {Structured
  micro/nano materials synthesized via electrospray: A review},}\ }\href
  {\doibase 10.1039/D0BM01313G} {\ \textbf {\bibinfo {volume} {8}},\ \bibinfo
  {pages} {5555--5573}}\BibitemShut {NoStop}%
\bibitem [{\citenamefont {Gamero-Castano}\ and\ \citenamefont
  {Hruby}()}]{Gamero-Castano2001}%
  \BibitemOpen
  \bibfield  {author} {\bibinfo {author} {\bibfnamefont {M.}~\bibnamefont
  {Gamero-Castano}}\ and\ \bibinfo {author} {\bibfnamefont {V.}~\bibnamefont
  {Hruby}},\ }\bibfield  {title} {\enquote {\bibinfo {title} {Electrospray as a
  {{Source}} of {{Nanoparticles}} for {{Efficient Colloid Thrusters}}},}\
  }\href {\doibase 10.2514/2.5858} {\ \textbf {\bibinfo {volume} {17}},\
  \bibinfo {pages} {977--987}}\BibitemShut {NoStop}%
\bibitem [{\citenamefont {Lozano}\ and\ \citenamefont
  {Martínez-Sánchez}({\natexlab{a}})}]{Lozano2005}%
  \BibitemOpen
  \bibfield  {author} {\bibinfo {author} {\bibfnamefont {P.}~\bibnamefont
  {Lozano}}\ and\ \bibinfo {author} {\bibfnamefont {M.}~\bibnamefont
  {Martínez-Sánchez}},\ }\bibfield  {title} {\enquote {\bibinfo {title}
  {Ionic liquid ion sources: Characterization of externally wetted emitters},}\
  }\href {\doibase 10.1016/j.jcis.2004.08.132} {\ \textbf {\bibinfo {volume}
  {282}},\ \bibinfo {pages} {415--421} ({\natexlab{a}})}\BibitemShut {NoStop}%
\bibitem [{\citenamefont {Gañán-Calvo}\ \emph {et~al.}()\citenamefont
  {Gañán-Calvo}, \citenamefont {López-Herrera}, \citenamefont {Herrada},
  \citenamefont {Ramos},\ and\ \citenamefont {Montanero}}]{Ganan-Calvo2018}%
  \BibitemOpen
  \bibfield  {author} {\bibinfo {author} {\bibfnamefont {A.~M.}\ \bibnamefont
  {Gañán-Calvo}}, \bibinfo {author} {\bibfnamefont {J.~M.}\ \bibnamefont
  {López-Herrera}}, \bibinfo {author} {\bibfnamefont {M.~A.}\ \bibnamefont
  {Herrada}}, \bibinfo {author} {\bibfnamefont {A.}~\bibnamefont {Ramos}}, \
  and\ \bibinfo {author} {\bibfnamefont {J.~M.}\ \bibnamefont {Montanero}},\
  }\bibfield  {title} {\enquote {\bibinfo {title} {Review on the physics of
  electrospray: {{From}} electrokinetics to the operating conditions of single
  and coaxial {{Taylor}} cone-jets, and {{AC}} electrospray},}\ }\href
  {\doibase 10.1016/j.jaerosci.2018.05.002} {\ \textbf {\bibinfo {volume}
  {125}},\ \bibinfo {pages} {32--56}}\BibitemShut {NoStop}%
\bibitem [{\citenamefont {Romero-Sanz}\ \emph {et~al.}()\citenamefont
  {Romero-Sanz}, \citenamefont {Bocanegra}, \citenamefont {Fernandez de~la
  Mora},\ and\ \citenamefont {Gamero-Castaño}}]{Romero-Sanz2003}%
  \BibitemOpen
  \bibfield  {author} {\bibinfo {author} {\bibfnamefont {I.}~\bibnamefont
  {Romero-Sanz}}, \bibinfo {author} {\bibfnamefont {R.}~\bibnamefont
  {Bocanegra}}, \bibinfo {author} {\bibfnamefont {J.}~\bibnamefont {Fernandez
  de~la Mora}}, \ and\ \bibinfo {author} {\bibfnamefont {M.}~\bibnamefont
  {Gamero-Castaño}},\ }\bibfield  {title} {\enquote {\bibinfo {title} {Source
  of heavy molecular ions based on {{Taylor}} cones of ionic liquids operating
  in the pure ion evaporation regime},}\ }\href {\doibase 10.1063/1.1598281} {\
  \textbf {\bibinfo {volume} {94}},\ \bibinfo {pages} {3599--3605}}\BibitemShut
  {NoStop}%
\bibitem [{\citenamefont {Gamero-Castaño}\ and\ \citenamefont
  {Mahadevan}()}]{Gamero-Castano2009b}%
  \BibitemOpen
  \bibfield  {author} {\bibinfo {author} {\bibfnamefont {M.}~\bibnamefont
  {Gamero-Castaño}}\ and\ \bibinfo {author} {\bibfnamefont {M.}~\bibnamefont
  {Mahadevan}},\ }\bibfield  {title} {\enquote {\bibinfo {title} {Sputtering of
  silicon by a beamlet of electrosprayed nanodroplets},}\ }\href {\doibase
  10.1016/j.apsusc.2009.06.018} {\ \textbf {\bibinfo {volume} {255}},\ \bibinfo
  {pages} {8556--8561}}\BibitemShut {NoStop}%
\bibitem [{\citenamefont {Cisquella-Serra}, \citenamefont
  {Galobardes-Esteban},\ and\ \citenamefont
  {Gamero-Castaño}()}]{Cisquella-Serra2022}%
  \BibitemOpen
  \bibfield  {author} {\bibinfo {author} {\bibfnamefont {A.}~\bibnamefont
  {Cisquella-Serra}}, \bibinfo {author} {\bibfnamefont {M.}~\bibnamefont
  {Galobardes-Esteban}}, \ and\ \bibinfo {author} {\bibfnamefont
  {M.}~\bibnamefont {Gamero-Castaño}},\ }\bibfield  {title} {\enquote
  {\bibinfo {title} {Scalable {{Microfabrication}} of {{Multi-Emitter Arrays}}
  in {{Silicon}} for a {{Compact Microfluidic Electrospray Propulsion
  System}}},}\ }\href {\doibase 10.1021/acsami.2c12716} {\ \textbf {\bibinfo
  {volume} {14}},\ \bibinfo {pages} {43527--43537}}\BibitemShut {NoStop}%
\bibitem [{\citenamefont {Krejci}\ and\ \citenamefont {Lozano}()}]{Krejci2016}%
  \BibitemOpen
  \bibfield  {author} {\bibinfo {author} {\bibfnamefont {D.}~\bibnamefont
  {Krejci}}\ and\ \bibinfo {author} {\bibfnamefont {P.}~\bibnamefont
  {Lozano}},\ }\bibfield  {title} {\enquote {\bibinfo {title} {{{SCALABLE IONIC
  LIQUID ELECTROSPRAY THRUSTERS FOR NANOSATELLITES}}},}\ }\href@noop {} {\ ,\
  \bibinfo {pages} {11}}\BibitemShut {NoStop}%
\bibitem [{\citenamefont {Magnani}\ and\ \citenamefont
  {Gamero-Castaño}({\natexlab{a}})}]{Magnani2024}%
  \BibitemOpen
  \bibfield  {author} {\bibinfo {author} {\bibfnamefont {M.}~\bibnamefont
  {Magnani}}\ and\ \bibinfo {author} {\bibfnamefont {M.}~\bibnamefont
  {Gamero-Castaño}},\ }\bibfield  {title} {\enquote {\bibinfo {title}
  {Analysis of self-heating in electrosprays operating in the cone-jet mode},}\
  }\href {\doibase 10.1017/jfm.2024.59} {\ \textbf {\bibinfo {volume} {980}},\
  \bibinfo {pages} {A40} ({\natexlab{a}})}\BibitemShut {NoStop}%
\bibitem [{\citenamefont {Beckey}\ and\ \citenamefont
  {Schulten}()}]{Beckey1975}%
  \BibitemOpen
  \bibfield  {author} {\bibinfo {author} {\bibfnamefont {H.~D.}\ \bibnamefont
  {Beckey}}\ and\ \bibinfo {author} {\bibfnamefont {H.-R.}\ \bibnamefont
  {Schulten}},\ }\bibfield  {title} {\enquote {\bibinfo {title} {Field
  {{Desorption Mass Spectrometry}}},}\ }\href {\doibase 10.1002/anie.197504031}
  {\ \textbf {\bibinfo {volume} {14}},\ \bibinfo {pages}
  {403--415}}\BibitemShut {NoStop}%
\bibitem [{\citenamefont {Gomer}()}]{Gomer1979}%
  \BibitemOpen
  \bibfield  {author} {\bibinfo {author} {\bibfnamefont {R.}~\bibnamefont
  {Gomer}},\ }\bibfield  {title} {\enquote {\bibinfo {title} {On the mechanism
  of liquid metal electron and ion sources},}\ }\href {\doibase
  10.1007/BF00930099} {\ \textbf {\bibinfo {volume} {19}},\ \bibinfo {pages}
  {365--375}}\BibitemShut {NoStop}%
\bibitem [{\citenamefont {Kelly}\ and\ \citenamefont {Miller}()}]{Kelly2007}%
  \BibitemOpen
  \bibfield  {author} {\bibinfo {author} {\bibfnamefont {T.~F.}\ \bibnamefont
  {Kelly}}\ and\ \bibinfo {author} {\bibfnamefont {M.~K.}\ \bibnamefont
  {Miller}},\ }\bibfield  {title} {\enquote {\bibinfo {title} {Atom probe
  tomography},}\ }\href {\doibase 10.1063/1.2709758} {\ \textbf {\bibinfo
  {volume} {78}},\ \bibinfo {pages} {031101}}\BibitemShut {NoStop}%
\bibitem [{\citenamefont {Caballero-Pérez}, \citenamefont
  {Galobardes-Esteban},\ and\ \citenamefont
  {Gamero-Castaño}()}]{Caballero-Perez}%
  \BibitemOpen
  \bibfield  {author} {\bibinfo {author} {\bibfnamefont {M.}~\bibnamefont
  {Caballero-Pérez}}, \bibinfo {author} {\bibfnamefont {M.}~\bibnamefont
  {Galobardes-Esteban}}, \ and\ \bibinfo {author} {\bibfnamefont
  {M.}~\bibnamefont {Gamero-Castaño}},\ }\bibfield  {title} {\enquote
  {\bibinfo {title} {High-{{Specific-Impulse Electrospray Propulsion}} with
  {{Small Capillary Emitters}}},}\ }\href {\doibase 10.2514/1.B40118} {\
  \textbf {\bibinfo {volume} {0}},\ \bibinfo {pages} {1--13}}\BibitemShut
  {NoStop}%
\bibitem [{\citenamefont {Gamero-Castaño}\ and\ \citenamefont
  {Magnani}({\natexlab{a}})}]{Gamero-Castano2019}%
  \BibitemOpen
  \bibfield  {author} {\bibinfo {author} {\bibfnamefont {M.}~\bibnamefont
  {Gamero-Castaño}}\ and\ \bibinfo {author} {\bibfnamefont {M.}~\bibnamefont
  {Magnani}},\ }\bibfield  {title} {\enquote {\bibinfo {title} {Numerical
  simulation of electrospraying in the cone-jet mode},}\ }\href {\doibase
  10.1017/jfm.2018.832} {\ \textbf {\bibinfo {volume} {859}},\ \bibinfo {pages}
  {247--267} ({\natexlab{a}})}\BibitemShut {NoStop}%
\bibitem [{\citenamefont {Ponce-Torres}\ \emph {et~al.}()\citenamefont
  {Ponce-Torres}, \citenamefont {Rebollo-Muñoz}, \citenamefont {Herrada},
  \citenamefont {Gañán-Calvo},\ and\ \citenamefont
  {Montanero}}]{Ponce-Torres2018}%
  \BibitemOpen
  \bibfield  {author} {\bibinfo {author} {\bibfnamefont {A.}~\bibnamefont
  {Ponce-Torres}}, \bibinfo {author} {\bibfnamefont {N.}~\bibnamefont
  {Rebollo-Muñoz}}, \bibinfo {author} {\bibfnamefont {M.~A.}\ \bibnamefont
  {Herrada}}, \bibinfo {author} {\bibfnamefont {A.~M.}\ \bibnamefont
  {Gañán-Calvo}}, \ and\ \bibinfo {author} {\bibfnamefont {J.~M.}\
  \bibnamefont {Montanero}},\ }\bibfield  {title} {\enquote {\bibinfo {title}
  {The steady cone-jet mode of electrospraying close to the minimum volume
  stability limit},}\ }\href {\doibase 10.1017/jfm.2018.737} {\ \textbf
  {\bibinfo {volume} {857}},\ \bibinfo {pages} {142--172}}\BibitemShut
  {NoStop}%
\bibitem [{\citenamefont {Gamero-Castaño}\ and\ \citenamefont
  {Magnani}({\natexlab{b}})}]{Gamero-Castano2019a}%
  \BibitemOpen
  \bibfield  {author} {\bibinfo {author} {\bibfnamefont {M.}~\bibnamefont
  {Gamero-Castaño}}\ and\ \bibinfo {author} {\bibfnamefont {M.}~\bibnamefont
  {Magnani}},\ }\bibfield  {title} {\enquote {\bibinfo {title} {The minimum
  flow rate of electrosprays in the cone-jet mode},}\ }\href {\doibase
  10.1017/jfm.2019.569} {\ \textbf {\bibinfo {volume} {876}},\ \bibinfo {pages}
  {553--572} ({\natexlab{b}})}\BibitemShut {NoStop}%
\bibitem [{\citenamefont {Gañán-Calvo}\ and\ \citenamefont
  {Montanero}()}]{Ganan-Calvo2009}%
  \BibitemOpen
  \bibfield  {author} {\bibinfo {author} {\bibfnamefont {A.~M.}\ \bibnamefont
  {Gañán-Calvo}}\ and\ \bibinfo {author} {\bibfnamefont {J.~M.}\ \bibnamefont
  {Montanero}},\ }\bibfield  {title} {\enquote {\bibinfo {title} {Revision of
  capillary cone-jet physics: {{Electrospray}} and flow focusing},}\ }\href
  {\doibase 10.1103/PhysRevE.79.066305} {\ \textbf {\bibinfo {volume} {79}},\
  \bibinfo {pages} {066305}}\BibitemShut {NoStop}%
\bibitem [{\citenamefont {Cloupeau}\ and\ \citenamefont
  {Prunet-Foch}()}]{Cloupeau1989}%
  \BibitemOpen
  \bibfield  {author} {\bibinfo {author} {\bibfnamefont {M.}~\bibnamefont
  {Cloupeau}}\ and\ \bibinfo {author} {\bibfnamefont {B.}~\bibnamefont
  {Prunet-Foch}},\ }\bibfield  {title} {\enquote {\bibinfo {title}
  {Electrostatic spraying of liquids in cone-jet mode},}\ }\href {\doibase
  10.1016/0304-3886(89)90081-8} {\ \textbf {\bibinfo {volume} {22}},\ \bibinfo
  {pages} {135--159}}\BibitemShut {NoStop}%
\bibitem [{\citenamefont {Gañán-Calvo}, \citenamefont {Rebollo-Muñoz},\ and\
  \citenamefont {Montanero}()}]{Ganan-Calvo2013}%
  \BibitemOpen
  \bibfield  {author} {\bibinfo {author} {\bibfnamefont {A.~M.}\ \bibnamefont
  {Gañán-Calvo}}, \bibinfo {author} {\bibfnamefont {N.}~\bibnamefont
  {Rebollo-Muñoz}}, \ and\ \bibinfo {author} {\bibfnamefont {J.~M.}\
  \bibnamefont {Montanero}},\ }\bibfield  {title} {\enquote {\bibinfo {title}
  {The minimum or natural rate of flow and droplet size ejected by {{Taylor}}
  cone–jets: Physical symmetries and scaling laws},}\ }\href {\doibase
  10.1088/1367-2630/15/3/033035} {\ \textbf {\bibinfo {volume} {15}},\ \bibinfo
  {pages} {033035}}\BibitemShut {NoStop}%
\bibitem [{\citenamefont {Higuera}()}]{Higuera2017}%
  \BibitemOpen
  \bibfield  {author} {\bibinfo {author} {\bibfnamefont {F.~J.}\ \bibnamefont
  {Higuera}},\ }\bibfield  {title} {\enquote {\bibinfo {title} {Qualitative
  analysis of the minimum flow rate of a cone-jet of a very polar liquid},}\
  }\href {\doibase 10.1017/jfm.2017.111} {\ \textbf {\bibinfo {volume} {816}},\
  \bibinfo {pages} {428--441}}\BibitemShut {NoStop}%
\bibitem [{\citenamefont {Perez-Lorenzo}\ and\ \citenamefont
  {family=Mora}()}]{Perez-Lorenzo2022}%
  \BibitemOpen
  \bibfield  {author} {\bibinfo {author} {\bibfnamefont {L.~J.}\ \bibnamefont
  {Perez-Lorenzo}}\ and\ \bibinfo {author} {\bibfnamefont {p.~l.~u.}\
  \bibnamefont {family=Mora}, \bibfnamefont {given=Juan~Fernandez}},\
  }\bibfield  {title} {\enquote {\bibinfo {title} {Probing electrically driven
  nanojets by energy and mass analysis in vacuo},}\ }\href {\doibase
  10.1017/jfm.2021.771} {\ \textbf {\bibinfo {volume} {931}},\ \bibinfo {pages}
  {A4}}\BibitemShut {NoStop}%
\bibitem [{\citenamefont {Magnani}, \citenamefont {Gamero-Castaño},\ and\
  \citenamefont {family=Mora}()}]{Magnani2025a}%
  \BibitemOpen
  \bibfield  {author} {\bibinfo {author} {\bibfnamefont {M.}~\bibnamefont
  {Magnani}}, \bibinfo {author} {\bibfnamefont {M.}~\bibnamefont
  {Gamero-Castaño}}, \ and\ \bibinfo {author} {\bibfnamefont {p.~l.~u.}\
  \bibnamefont {family=Mora}, \bibfnamefont {given=Juan~Fernández}},\
  }\bibfield  {title} {\enquote {\bibinfo {title} {Determination of the
  characteristic length of electrosprays operating in the cone-jet mode},}\
  }\href {\doibase 10.1017/jfm.2025.10529} {\ \textbf {\bibinfo {volume}
  {1018}},\ \bibinfo {pages} {A5}}\BibitemShut {NoStop}%
\bibitem [{\citenamefont
  {Gamero-Castaño}({\natexlab{a}})}]{Gamero-Castano2010}%
  \BibitemOpen
  \bibfield  {author} {\bibinfo {author} {\bibfnamefont {M.}~\bibnamefont
  {Gamero-Castaño}},\ }\bibfield  {title} {\enquote {\bibinfo {title} {Energy
  dissipation in electrosprays and the geometric scaling of the transition
  region of cone–jets},}\ }\href {\doibase 10.1017/S0022112010003423} {\
  \textbf {\bibinfo {volume} {662}},\ \bibinfo {pages} {493--513}
  ({\natexlab{a}})}\BibitemShut {NoStop}%
\bibitem [{\citenamefont {Magnani}, \citenamefont {Caballero-Pérez},\ and\
  \citenamefont {Gamero-Castaño}()}]{Magnani2025}%
  \BibitemOpen
  \bibfield  {author} {\bibinfo {author} {\bibfnamefont {M.}~\bibnamefont
  {Magnani}}, \bibinfo {author} {\bibfnamefont {M.}~\bibnamefont
  {Caballero-Pérez}}, \ and\ \bibinfo {author} {\bibfnamefont
  {M.}~\bibnamefont {Gamero-Castaño}},\ }\bibfield  {title} {\enquote
  {\bibinfo {title} {Modelling of electrosprays of ionic liquids including
  dissipation and self-heating},}\ }\href {\doibase 10.1017/jfm.2025.10549} {\
  \textbf {\bibinfo {volume} {1018}},\ \bibinfo {pages} {A39}}\BibitemShut
  {NoStop}%
\bibitem [{\citenamefont {family=Mitropoulos}()}]{Mitropoulos2008}%
  \BibitemOpen
  \bibfield  {author} {\bibinfo {author} {\bibfnamefont {g.-i.}\ \bibnamefont
  {family=Mitropoulos}, \bibfnamefont {given=A.~Ch.}},\ }\bibfield  {title}
  {\enquote {\bibinfo {title} {The {{Kelvin}} equation},}\ }\href {\doibase
  10.1016/j.jcis.2007.10.001} {\ \textbf {\bibinfo {volume} {317}},\ \bibinfo
  {pages} {643--648}}\BibitemShut {NoStop}%
\bibitem [{Note1()}]{Note1}%
  \BibitemOpen
  \bibinfo {note} {BMI: 1-butyl-3-methylimidazolium, DCA: dicyanamide, BF$_4$:
  tetrafluoroborate, FAP: tris(pentafluoroethyl)trifluorophosphate}\BibitemShut
  {NoStop}%
\bibitem [{\citenamefont {Miller}\ \emph {et~al.}()\citenamefont {Miller},
  \citenamefont {Ulibarri-Sanchez}, \citenamefont {Prince},\ and\ \citenamefont
  {Bemish}}]{Miller2021}%
  \BibitemOpen
  \bibfield  {author} {\bibinfo {author} {\bibfnamefont {S.~W.}\ \bibnamefont
  {Miller}}, \bibinfo {author} {\bibfnamefont {J.~R.}\ \bibnamefont
  {Ulibarri-Sanchez}}, \bibinfo {author} {\bibfnamefont {B.~D.}\ \bibnamefont
  {Prince}}, \ and\ \bibinfo {author} {\bibfnamefont {R.~J.}\ \bibnamefont
  {Bemish}},\ }\bibfield  {title} {\enquote {\bibinfo {title} {Capillary ionic
  liquid electrospray: Beam compositional analysis by orthogonal time-of-flight
  mass spectrometry},}\ }\href {\doibase 10.1017/jfm.2021.783} {\ \textbf
  {\bibinfo {volume} {928}},\ \bibinfo {pages} {A12}}\BibitemShut {NoStop}%
\bibitem [{\citenamefont {Gamero-Castaño}\ and\ \citenamefont
  {Cisquella-Serra}()}]{Gamero-Castano2021}%
  \BibitemOpen
  \bibfield  {author} {\bibinfo {author} {\bibfnamefont {M.}~\bibnamefont
  {Gamero-Castaño}}\ and\ \bibinfo {author} {\bibfnamefont {A.}~\bibnamefont
  {Cisquella-Serra}},\ }\bibfield  {title} {\enquote {\bibinfo {title}
  {Electrosprays of highly conducting liquids: {{A}} study of droplet and ion
  emission based on retarding potential and time-of-flight spectrometry},}\
  }\href {\doibase 10.1103/PhysRevFluids.6.013701} {\ \textbf {\bibinfo
  {volume} {6}},\ \bibinfo {pages} {013701}}\BibitemShut {NoStop}%
\bibitem [{\citenamefont {Misra}\ and\ \citenamefont
  {Gamero-Castaño}({\natexlab{a}})}]{Misra2022}%
  \BibitemOpen
  \bibfield  {author} {\bibinfo {author} {\bibfnamefont {K.}~\bibnamefont
  {Misra}}\ and\ \bibinfo {author} {\bibfnamefont {M.}~\bibnamefont
  {Gamero-Castaño}},\ }\bibfield  {title} {\enquote {\bibinfo {title}
  {Leaky-dielectric phase field model for the axisymmetric breakup of an
  electrified jet},}\ }\href {\doibase 10.1103/PhysRevFluids.7.064004} {\
  \textbf {\bibinfo {volume} {7}},\ \bibinfo {pages} {064004}
  ({\natexlab{a}})}\BibitemShut {NoStop}%
\bibitem [{\citenamefont {Yang}\ \emph {et~al.}()\citenamefont {Yang},
  \citenamefont {Duan}, \citenamefont {Li},\ and\ \citenamefont
  {Deng}}]{Yang2014}%
  \BibitemOpen
  \bibfield  {author} {\bibinfo {author} {\bibfnamefont {W.}~\bibnamefont
  {Yang}}, \bibinfo {author} {\bibfnamefont {H.}~\bibnamefont {Duan}}, \bibinfo
  {author} {\bibfnamefont {C.}~\bibnamefont {Li}}, \ and\ \bibinfo {author}
  {\bibfnamefont {W.}~\bibnamefont {Deng}},\ }\bibfield  {title} {\enquote
  {\bibinfo {title} {Crossover of {{Varicose}} and {{Whipping Instabilities}}
  in {{Electrified Microjets}}},}\ }\href {\doibase
  10.1103/PhysRevLett.112.054501} {\ \textbf {\bibinfo {volume} {112}},\
  \bibinfo {pages} {054501}}\BibitemShut {NoStop}%
\bibitem [{\citenamefont
  {Gamero-Castaño}({\natexlab{b}})}]{Gamero-Castano2009}%
  \BibitemOpen
  \bibfield  {author} {\bibinfo {author} {\bibfnamefont {M.}~\bibnamefont
  {Gamero-Castaño}},\ }\bibfield  {title} {\enquote {\bibinfo {title}
  {Retarding potential and induction charge detectors in tandem for measuring
  the charge and mass of nanodroplets},}\ }\href {\doibase 10.1063/1.3128730}
  {\ \textbf {\bibinfo {volume} {80}},\ \bibinfo {pages} {053301}
  ({\natexlab{b}})}\BibitemShut {NoStop}%
\bibitem [{\citenamefont {Gamero-Castaño}\ and\ \citenamefont
  {Hruby}()}]{Gamero-Castano2002}%
  \BibitemOpen
  \bibfield  {author} {\bibinfo {author} {\bibfnamefont {M.}~\bibnamefont
  {Gamero-Castaño}}\ and\ \bibinfo {author} {\bibfnamefont {V.}~\bibnamefont
  {Hruby}},\ }\bibfield  {title} {\enquote {\bibinfo {title} {Electric
  measurements of charged sprays emitted by cone-jets},}\ }\href {\doibase
  10.1017/S002211200200798X} {\ \textbf {\bibinfo {volume} {459}},\ \bibinfo
  {pages} {245--276}}\BibitemShut {NoStop}%
\bibitem [{\citenamefont {Schweizer}\ and\ \citenamefont
  {Hanson}()}]{Schweizer1971}%
  \BibitemOpen
  \bibfield  {author} {\bibinfo {author} {\bibfnamefont {J.~W.}\ \bibnamefont
  {Schweizer}}\ and\ \bibinfo {author} {\bibfnamefont {D.~N.}\ \bibnamefont
  {Hanson}},\ }\bibfield  {title} {\enquote {\bibinfo {title} {Stability limit
  of charged drops},}\ }\href {\doibase 10.1016/0021-9797(71)90141-X} {\
  \textbf {\bibinfo {volume} {35}},\ \bibinfo {pages} {417--423}}\BibitemShut
  {NoStop}%
\bibitem [{\citenamefont {Abbas}\ and\ \citenamefont {Latham}()}]{Abbas1967}%
  \BibitemOpen
  \bibfield  {author} {\bibinfo {author} {\bibfnamefont {M.~A.}\ \bibnamefont
  {Abbas}}\ and\ \bibinfo {author} {\bibfnamefont {J.}~\bibnamefont {Latham}},\
  }\bibfield  {title} {\enquote {\bibinfo {title} {The instability of
  evaporating charged drops},}\ }\href {\doibase 10.1017/S0022112067001685} {\
  \textbf {\bibinfo {volume} {30}},\ \bibinfo {pages} {663--670}}\BibitemShut
  {NoStop}%
\bibitem [{\citenamefont {Richardson}, \citenamefont {Pigg},\ and\
  \citenamefont {Hightower}()}]{Richardson1997}%
  \BibitemOpen
  \bibfield  {author} {\bibinfo {author} {\bibfnamefont {C.~B.}\ \bibnamefont
  {Richardson}}, \bibinfo {author} {\bibfnamefont {A.~L.}\ \bibnamefont
  {Pigg}}, \ and\ \bibinfo {author} {\bibfnamefont {R.~L.}\ \bibnamefont
  {Hightower}},\ }\bibfield  {title} {\enquote {\bibinfo {title} {On the
  stability limit of charged droplets},}\ }\href {\doibase
  10.1098/rspa.1989.0031} {\ \textbf {\bibinfo {volume} {422}},\ \bibinfo
  {pages} {319--328}}\BibitemShut {NoStop}%
\bibitem [{\citenamefont {Singh}\ \emph {et~al.}()\citenamefont {Singh},
  \citenamefont {Gawande}, \citenamefont {Mayya},\ and\ \citenamefont
  {Thaokar}}]{Singh2021}%
  \BibitemOpen
  \bibfield  {author} {\bibinfo {author} {\bibfnamefont {M.}~\bibnamefont
  {Singh}}, \bibinfo {author} {\bibfnamefont {N.}~\bibnamefont {Gawande}},
  \bibinfo {author} {\bibfnamefont {Y.~S.}\ \bibnamefont {Mayya}}, \ and\
  \bibinfo {author} {\bibfnamefont {R.}~\bibnamefont {Thaokar}},\ }\bibfield
  {title} {\enquote {\bibinfo {title} {Subcritical asymmetric {{Rayleigh}}
  breakup of a charged drop induced by finite amplitude perturbations in a
  quadrupole trap},}\ }\href {\doibase 10.1103/PhysRevE.103.053111} {\ \textbf
  {\bibinfo {volume} {103}},\ \bibinfo {pages} {053111}}\BibitemShut {NoStop}%
\bibitem [{\citenamefont {Misra}\ and\ \citenamefont
  {Gamero-Castaño}({\natexlab{b}})}]{Misra2023}%
  \BibitemOpen
  \bibfield  {author} {\bibinfo {author} {\bibfnamefont {K.}~\bibnamefont
  {Misra}}\ and\ \bibinfo {author} {\bibfnamefont {M.}~\bibnamefont
  {Gamero-Castaño}},\ }\bibfield  {title} {\enquote {\bibinfo {title} {Ion
  emission from nanodroplets undergoing {{Coulomb}} explosions: A continuum
  numerical study},}\ }\href {\doibase 10.1017/jfm.2023.107} {\ \textbf
  {\bibinfo {volume} {958}},\ \bibinfo {pages} {A32}
  ({\natexlab{b}})}\BibitemShut {NoStop}%
\bibitem [{\citenamefont
  {Gamero-Castaño}({\natexlab{c}})}]{Gamero-Castano2008}%
  \BibitemOpen
  \bibfield  {author} {\bibinfo {author} {\bibfnamefont {M.}~\bibnamefont
  {Gamero-Castaño}},\ }\bibfield  {title} {\enquote {\bibinfo {title}
  {Characterization of the electrosprays of 1-ethyl-3-methylimidazolium
  bis(trifluoromethylsulfonyl) imide in vacuum},}\ }\href {\doibase
  10.1063/1.2899658} {\ \textbf {\bibinfo {volume} {20}},\ \bibinfo {pages}
  {032103} ({\natexlab{c}})}\BibitemShut {NoStop}%
\bibitem [{\citenamefont {Lyne}, \citenamefont {Liu},\ and\ \citenamefont
  {Rovey}()}]{Lyne2023}%
  \BibitemOpen
  \bibfield  {author} {\bibinfo {author} {\bibfnamefont {C.~T.}\ \bibnamefont
  {Lyne}}, \bibinfo {author} {\bibfnamefont {M.~F.}\ \bibnamefont {Liu}}, \
  and\ \bibinfo {author} {\bibfnamefont {J.~L.}\ \bibnamefont {Rovey}},\
  }\bibfield  {title} {\enquote {\bibinfo {title} {A simple retarding-potential
  time-of-flight mass spectrometer for electrospray propulsion diagnostics},}\
  }\href {\doibase 10.1007/s44205-023-00045-y} {\ \textbf {\bibinfo {volume}
  {2}},\ \bibinfo {pages} {13}}\BibitemShut {NoStop}%
\bibitem [{\citenamefont {Iribarne}\ and\ \citenamefont
  {Thomson}()}]{Iribarne1976}%
  \BibitemOpen
  \bibfield  {author} {\bibinfo {author} {\bibfnamefont {J.~V.}\ \bibnamefont
  {Iribarne}}\ and\ \bibinfo {author} {\bibfnamefont {B.~A.}\ \bibnamefont
  {Thomson}},\ }\bibfield  {title} {\enquote {\bibinfo {title} {On the
  evaporation of small ions from charged droplets},}\ }\href {\doibase
  10.1063/1.432536} {\ \textbf {\bibinfo {volume} {64}},\ \bibinfo {pages}
  {2287--2294}}\BibitemShut {NoStop}%
\bibitem [{\citenamefont {Thomson}\ and\ \citenamefont
  {Iribarne}()}]{Thomson1979}%
  \BibitemOpen
  \bibfield  {author} {\bibinfo {author} {\bibfnamefont {B.~A.}\ \bibnamefont
  {Thomson}}\ and\ \bibinfo {author} {\bibfnamefont {J.~V.}\ \bibnamefont
  {Iribarne}},\ }\bibfield  {title} {\enquote {\bibinfo {title} {Field induced
  ion evaporation from liquid surfaces at atmospheric pressure},}\ }\href
  {\doibase 10.1063/1.438198} {\ \textbf {\bibinfo {volume} {71}},\ \bibinfo
  {pages} {4451--4463}}\BibitemShut {NoStop}%
\bibitem [{\citenamefont {Magnani}\ and\ \citenamefont
  {Gamero-Castaño}({\natexlab{b}})}]{Magnani2022}%
  \BibitemOpen
  \bibfield  {author} {\bibinfo {author} {\bibfnamefont {M.}~\bibnamefont
  {Magnani}}\ and\ \bibinfo {author} {\bibfnamefont {M.}~\bibnamefont
  {Gamero-Castaño}},\ }\bibfield  {title} {\enquote {\bibinfo {title} {Energy
  barrier for ion field emission from a dielectric liquid sphere},}\ }\href
  {\doibase 10.1103/PhysRevE.105.054802} {\ \textbf {\bibinfo {volume} {105}},\
  \bibinfo {pages} {054802} ({\natexlab{b}})}\BibitemShut {NoStop}%
\bibitem [{\citenamefont {Labowsky}, \citenamefont {Fenn},\ and\ \citenamefont
  {Fernandez de~la Mora}()}]{Labowsky2000}%
  \BibitemOpen
  \bibfield  {author} {\bibinfo {author} {\bibfnamefont {M.}~\bibnamefont
  {Labowsky}}, \bibinfo {author} {\bibfnamefont {J.~B.}\ \bibnamefont {Fenn}},
  \ and\ \bibinfo {author} {\bibfnamefont {J.}~\bibnamefont {Fernandez de~la
  Mora}},\ }\bibfield  {title} {\enquote {\bibinfo {title} {A continuum model
  for ion evaporation from a drop: Effect of curvature and charge on ion
  solvation energy},}\ }\href {\doibase 10.1016/S0003-2670(99)00595-4} {\
  \textbf {\bibinfo {volume} {406}},\ \bibinfo {pages} {105--118}}\BibitemShut
  {NoStop}%
\bibitem [{\citenamefont {Yu}\ and\ \citenamefont {Chen}()}]{Yu2021}%
  \BibitemOpen
  \bibfield  {author} {\bibinfo {author} {\bibfnamefont {Y.}~\bibnamefont
  {Yu}}\ and\ \bibinfo {author} {\bibfnamefont {Y.}~\bibnamefont {Chen}},\
  }\bibfield  {title} {\enquote {\bibinfo {title} {Density {{Prediction}} of
  {{Ionic Liquids}} at {{Different Temperatures Using}} the {{Average Free
  Volume Model}}},}\ }\href {\doibase 10.1021/acsomega.1c00547} {\ \textbf
  {\bibinfo {volume} {6}},\ \bibinfo {pages} {14869--14874}}\BibitemShut
  {NoStop}%
\bibitem [{\citenamefont
  {Gamero-Castaño}({\natexlab{d}})}]{Gamero-Castano2010a}%
  \BibitemOpen
  \bibfield  {author} {\bibinfo {author} {\bibfnamefont {M.}~\bibnamefont
  {Gamero-Castaño}},\ }\bibfield  {title} {\enquote {\bibinfo {title} {Energy
  dissipation in electrosprays and the geometric scaling of the transition
  region of cone–jets},}\ }\href {\doibase 10.1017/S0022112010003423} {\
  \textbf {\bibinfo {volume} {662}},\ \bibinfo {pages} {493--513}
  ({\natexlab{d}})}\BibitemShut {NoStop}%
\bibitem [{\citenamefont {Feng}\ \emph {et~al.}()\citenamefont {Feng},
  \citenamefont {Chen}, \citenamefont {Bi}, \citenamefont {Goodwin},
  \citenamefont {Postnikov}, \citenamefont {Brilliantov}, \citenamefont
  {Urbakh},\ and\ \citenamefont {Kornyshev}}]{Feng2019}%
  \BibitemOpen
  \bibfield  {author} {\bibinfo {author} {\bibfnamefont {G.}~\bibnamefont
  {Feng}}, \bibinfo {author} {\bibfnamefont {M.}~\bibnamefont {Chen}}, \bibinfo
  {author} {\bibfnamefont {S.}~\bibnamefont {Bi}}, \bibinfo {author}
  {\bibfnamefont {Z.~A.~H.}\ \bibnamefont {Goodwin}}, \bibinfo {author}
  {\bibfnamefont {E.~B.}\ \bibnamefont {Postnikov}}, \bibinfo {author}
  {\bibfnamefont {N.}~\bibnamefont {Brilliantov}}, \bibinfo {author}
  {\bibfnamefont {M.}~\bibnamefont {Urbakh}}, \ and\ \bibinfo {author}
  {\bibfnamefont {A.~A.}\ \bibnamefont {Kornyshev}},\ }\bibfield  {title}
  {\enquote {\bibinfo {title} {Free and {{Bound States}} of {{Ions}} in {{Ionic
  Liquids}}, {{Conductivity}}, and {{Underscreening Paradox}}},}\ }\href
  {\doibase 10.1103/PhysRevX.9.021024} {\ \textbf {\bibinfo {volume} {9}},\
  \bibinfo {pages} {021024}}\BibitemShut {NoStop}%
\bibitem [{\citenamefont {MacFarlane}\ \emph {et~al.}()\citenamefont
  {MacFarlane}, \citenamefont {Forsyth}, \citenamefont {Izgorodina},
  \citenamefont {Abbott}, \citenamefont {Annat},\ and\ \citenamefont
  {Fraser}}]{MacFarlane2009}%
  \BibitemOpen
  \bibfield  {author} {\bibinfo {author} {\bibfnamefont {D.~R.}\ \bibnamefont
  {MacFarlane}}, \bibinfo {author} {\bibfnamefont {M.}~\bibnamefont {Forsyth}},
  \bibinfo {author} {\bibfnamefont {E.~I.}\ \bibnamefont {Izgorodina}},
  \bibinfo {author} {\bibfnamefont {A.~P.}\ \bibnamefont {Abbott}}, \bibinfo
  {author} {\bibfnamefont {G.}~\bibnamefont {Annat}}, \ and\ \bibinfo {author}
  {\bibfnamefont {K.}~\bibnamefont {Fraser}},\ }\bibfield  {title} {\enquote
  {\bibinfo {title} {On the concept of ionicity in ionic liquids},}\ }\href
  {\doibase 10.1039/B900201D} {\ \textbf {\bibinfo {volume} {11}},\ \bibinfo
  {pages} {4962--4967}}\BibitemShut {NoStop}%
\bibitem [{\citenamefont {Nordness}\ and\ \citenamefont
  {Brennecke}()}]{Nordness2020}%
  \BibitemOpen
  \bibfield  {author} {\bibinfo {author} {\bibfnamefont {O.}~\bibnamefont
  {Nordness}}\ and\ \bibinfo {author} {\bibfnamefont {J.~F.}\ \bibnamefont
  {Brennecke}},\ }\bibfield  {title} {\enquote {\bibinfo {title} {Ion
  {{Dissociation}} in {{Ionic Liquids}} and {{Ionic Liquid Solutions}}},}\
  }\href {\doibase 10.1021/acs.chemrev.0c00373} {\ \textbf {\bibinfo {volume}
  {120}},\ \bibinfo {pages} {12873--12902}}\BibitemShut {NoStop}%
\bibitem [{\citenamefont {C.~Kroon}\ \emph {et~al.}()\citenamefont {C.~Kroon},
  \citenamefont {Buijs}, \citenamefont {J.~Peters},\ and\ \citenamefont
  {Witkamp}}]{C.Kroon2006}%
  \BibitemOpen
  \bibfield  {author} {\bibinfo {author} {\bibfnamefont {M.}~\bibnamefont
  {C.~Kroon}}, \bibinfo {author} {\bibfnamefont {W.}~\bibnamefont {Buijs}},
  \bibinfo {author} {\bibfnamefont {C.}~\bibnamefont {J.~Peters}}, \ and\
  \bibinfo {author} {\bibfnamefont {G.-J.}\ \bibnamefont {Witkamp}},\
  }\bibfield  {title} {\enquote {\bibinfo {title} {Decomposition of ionic
  liquids in electrochemical processing},}\ }\href {\doibase 10.1039/B512724F}
  {\ \textbf {\bibinfo {volume} {8}},\ \bibinfo {pages} {241--245}}\BibitemShut
  {NoStop}%
\bibitem [{\citenamefont {Natisin}\ \emph {et~al.}()\citenamefont {Natisin},
  \citenamefont {Zamora}, \citenamefont {Holley}, \citenamefont {Ivan~Arnold},
  \citenamefont {McGehee}, \citenamefont {Holmes},\ and\ \citenamefont
  {Eckhardt}}]{Natisin2021}%
  \BibitemOpen
  \bibfield  {author} {\bibinfo {author} {\bibfnamefont {M.~R.}\ \bibnamefont
  {Natisin}}, \bibinfo {author} {\bibfnamefont {H.~L.}\ \bibnamefont {Zamora}},
  \bibinfo {author} {\bibfnamefont {Z.~A.}\ \bibnamefont {Holley}}, \bibinfo
  {author} {\bibfnamefont {N.}~\bibnamefont {Ivan~Arnold}}, \bibinfo {author}
  {\bibfnamefont {W.~A.}\ \bibnamefont {McGehee}}, \bibinfo {author}
  {\bibfnamefont {M.~R.}\ \bibnamefont {Holmes}}, \ and\ \bibinfo {author}
  {\bibfnamefont {D.}~\bibnamefont {Eckhardt}},\ }\bibfield  {title} {\enquote
  {\bibinfo {title} {Efficiency {{Mechanisms}} in {{Porous-Media Electrospray
  Thrusters}}},}\ }\href {\doibase 10.2514/1.B38160} {\ \textbf {\bibinfo
  {volume} {37}},\ \bibinfo {pages} {650--659}}\BibitemShut {NoStop}%
\bibitem [{\citenamefont {Huang}\ \emph {et~al.}()\citenamefont {Huang},
  \citenamefont {Li}, \citenamefont {Li}, \citenamefont {Si}, \citenamefont
  {Xiong},\ and\ \citenamefont {Fan}}]{Huang2021}%
  \BibitemOpen
  \bibfield  {author} {\bibinfo {author} {\bibfnamefont {C.}~\bibnamefont
  {Huang}}, \bibinfo {author} {\bibfnamefont {J.}~\bibnamefont {Li}}, \bibinfo
  {author} {\bibfnamefont {M.}~\bibnamefont {Li}}, \bibinfo {author}
  {\bibfnamefont {T.}~\bibnamefont {Si}}, \bibinfo {author} {\bibfnamefont
  {C.}~\bibnamefont {Xiong}}, \ and\ \bibinfo {author} {\bibfnamefont
  {W.}~\bibnamefont {Fan}},\ }\bibfield  {title} {\enquote {\bibinfo {title}
  {Experimental investigation on current modes of ionic liquid electrospray
  from a coned porous emitter},}\ }\href {\doibase
  10.1016/j.actaastro.2021.03.014} {\ \textbf {\bibinfo {volume} {183}},\
  \bibinfo {pages} {286--299}}\BibitemShut {NoStop}%
\bibitem [{\citenamefont {De~Saavedra}\ \emph {et~al.}()\citenamefont
  {De~Saavedra}, \citenamefont {Villegas-Prados}, \citenamefont {Wijnen},\ and\
  \citenamefont {Cruz}}]{DeSaavedra2025}%
  \BibitemOpen
  \bibfield  {author} {\bibinfo {author} {\bibfnamefont {B.}~\bibnamefont
  {De~Saavedra}}, \bibinfo {author} {\bibfnamefont {D.}~\bibnamefont
  {Villegas-Prados}}, \bibinfo {author} {\bibfnamefont {M.}~\bibnamefont
  {Wijnen}}, \ and\ \bibinfo {author} {\bibfnamefont {J.}~\bibnamefont
  {Cruz}},\ }\bibfield  {title} {\enquote {\bibinfo {title} {Comparison between
  direct and indirect measurements for externally wetted electrospray thruster
  with adapted analytical balance},}\ }\href {\doibase
  10.1016/j.actaastro.2025.02.044} {\ \textbf {\bibinfo {volume} {232}},\
  \bibinfo {pages} {283--295}}\BibitemShut {NoStop}%
\bibitem [{\citenamefont {Demmons}\ \emph {et~al.}()\citenamefont {Demmons},
  \citenamefont {Wood}, \citenamefont {Margousian}, \citenamefont {Knott},\
  and\ \citenamefont {Fedkiw}}]{Demmons}%
  \BibitemOpen
  \bibfield  {author} {\bibinfo {author} {\bibfnamefont {N.~R.}\ \bibnamefont
  {Demmons}}, \bibinfo {author} {\bibfnamefont {Z.~D.}\ \bibnamefont {Wood}},
  \bibinfo {author} {\bibfnamefont {A.}~\bibnamefont {Margousian}}, \bibinfo
  {author} {\bibfnamefont {J.}~\bibnamefont {Knott}}, \ and\ \bibinfo {author}
  {\bibfnamefont {T.}~\bibnamefont {Fedkiw}},\ }\bibfield  {title} {\enquote
  {\bibinfo {title} {Electrospray {{Attitude Control System Flight
  Preparation}}},}\ }in\ \href {\doibase 10.2514/6.2022-0039} {\emph {\bibinfo
  {booktitle} {{{AIAA SCITECH}} 2022 {{Forum}}}}}\ (\bibinfo  {publisher}
  {{American Institute of Aeronautics and Astronautics}})\BibitemShut {NoStop}%
\bibitem [{\citenamefont {Galobardes-Esteban}\ \emph {et~al.}()\citenamefont
  {Galobardes-Esteban}, \citenamefont {Cisquella-Serra}, \citenamefont
  {Caballero-Pérez},\ and\ \citenamefont
  {Gamero-Castaño}}]{Galobardes-Esteban2026}%
  \BibitemOpen
  \bibfield  {author} {\bibinfo {author} {\bibfnamefont {M.}~\bibnamefont
  {Galobardes-Esteban}}, \bibinfo {author} {\bibfnamefont {A.}~\bibnamefont
  {Cisquella-Serra}}, \bibinfo {author} {\bibfnamefont {M.}~\bibnamefont
  {Caballero-Pérez}}, \ and\ \bibinfo {author} {\bibfnamefont
  {M.}~\bibnamefont {Gamero-Castaño}},\ }\bibfield  {title} {\enquote
  {\bibinfo {title} {Variable {{Specific Impulse Electrospray Thruster}} for
  {{SmallSat}}},}\ }\href@noop {} {\ }\BibitemShut {NoStop}%
\bibitem [{\citenamefont {Corrado}\ \emph {et~al.}()\citenamefont {Corrado},
  \citenamefont {Wangari}, \citenamefont {Finch}, \citenamefont {Parameswaran},
  \citenamefont {Cook}, \citenamefont {Holihan}, \citenamefont {Mathews},
  \citenamefont {Kedzierski}, \citenamefont {Racz}, \citenamefont {Smith},\
  and\ \citenamefont {Lozano}}]{Corrado2024}%
  \BibitemOpen
  \bibfield  {author} {\bibinfo {author} {\bibfnamefont {M.~N.}\ \bibnamefont
  {Corrado}}, \bibinfo {author} {\bibfnamefont {C.}~\bibnamefont {Wangari}},
  \bibinfo {author} {\bibfnamefont {M.~R.}\ \bibnamefont {Finch}}, \bibinfo
  {author} {\bibfnamefont {L.}~\bibnamefont {Parameswaran}}, \bibinfo {author}
  {\bibfnamefont {M.}~\bibnamefont {Cook}}, \bibinfo {author} {\bibfnamefont
  {E.~C.}\ \bibnamefont {Holihan}}, \bibinfo {author} {\bibfnamefont
  {R.}~\bibnamefont {Mathews}}, \bibinfo {author} {\bibfnamefont
  {J.}~\bibnamefont {Kedzierski}}, \bibinfo {author} {\bibfnamefont {L.~M.}\
  \bibnamefont {Racz}}, \bibinfo {author} {\bibfnamefont {M.~A.}\ \bibnamefont
  {Smith}}, \ and\ \bibinfo {author} {\bibfnamefont {P.~C.}\ \bibnamefont
  {Lozano}},\ }\bibfield  {title} {\enquote {\bibinfo {title} {Performance
  {{Characterization}} of {{High-Density Electrospray Thrusters}}},}\ }\href
  {https://dspace.mit.edu/handle/1721.1/155454} {\ }\BibitemShut {NoStop}%
\bibitem [{\citenamefont {Krejci}\ \emph {et~al.}()\citenamefont {Krejci},
  \citenamefont {Mier-Hicks}, \citenamefont {Thomas}, \citenamefont {Haag},\
  and\ \citenamefont {Lozano}}]{Krejci2017}%
  \BibitemOpen
  \bibfield  {author} {\bibinfo {author} {\bibfnamefont {D.}~\bibnamefont
  {Krejci}}, \bibinfo {author} {\bibfnamefont {F.}~\bibnamefont {Mier-Hicks}},
  \bibinfo {author} {\bibfnamefont {R.}~\bibnamefont {Thomas}}, \bibinfo
  {author} {\bibfnamefont {T.}~\bibnamefont {Haag}}, \ and\ \bibinfo {author}
  {\bibfnamefont {P.}~\bibnamefont {Lozano}},\ }\bibfield  {title} {\enquote
  {\bibinfo {title} {Emission {{Characteristics}} of {{Passively Fed
  Electrospray Microthrusters}} with {{Propellant Reservoirs}}},}\ }\href
  {\doibase 10.2514/1.A33531} {\ \textbf {\bibinfo {volume} {54}},\ \bibinfo
  {pages} {447--458}}\BibitemShut {NoStop}%
\bibitem [{\citenamefont {Coffman}, \citenamefont {Martínez-Sánchez},\ and\
  \citenamefont {Lozano}()}]{Coffman2019}%
  \BibitemOpen
  \bibfield  {author} {\bibinfo {author} {\bibfnamefont {C.~S.}\ \bibnamefont
  {Coffman}}, \bibinfo {author} {\bibfnamefont {M.}~\bibnamefont
  {Martínez-Sánchez}}, \ and\ \bibinfo {author} {\bibfnamefont {P.~C.}\
  \bibnamefont {Lozano}},\ }\bibfield  {title} {\enquote {\bibinfo {title}
  {Electrohydrodynamics of an ionic liquid meniscus during evaporation of ions
  in a regime of high electric field},}\ }\href {\doibase
  10.1103/PhysRevE.99.063108} {\ \textbf {\bibinfo {volume} {99}},\ \bibinfo
  {pages} {063108}}\BibitemShut {NoStop}%
\bibitem [{\citenamefont {Magnani}\ and\ \citenamefont
  {Gamero-Castaño}({\natexlab{c}})}]{Magnani2023}%
  \BibitemOpen
  \bibfield  {author} {\bibinfo {author} {\bibfnamefont {M.}~\bibnamefont
  {Magnani}}\ and\ \bibinfo {author} {\bibfnamefont {M.}~\bibnamefont
  {Gamero-Castaño}},\ }\bibfield  {title} {\enquote {\bibinfo {title}
  {Modelling and scaling laws of the ion emission regime in {{Taylor}}
  cones},}\ }\href {\doibase 10.1017/jfm.2023.717} {\ \textbf {\bibinfo
  {volume} {972}},\ \bibinfo {pages} {A34} ({\natexlab{c}})}\BibitemShut
  {NoStop}%
\bibitem [{\citenamefont {Lozano}\ and\ \citenamefont
  {Martinez-Sanchez}()}]{Lozano2005a}%
  \BibitemOpen
  \bibfield  {author} {\bibinfo {author} {\bibfnamefont {P.}~\bibnamefont
  {Lozano}}\ and\ \bibinfo {author} {\bibfnamefont {M.}~\bibnamefont
  {Martinez-Sanchez}},\ }\bibfield  {title} {\enquote {\bibinfo {title}
  {Efficiency {{Estimation}} of {{EMI-BF4 Ionic Liquid Electrospray
  Thrusters}}},}\ }in\ \href {\doibase 10.2514/6.2005-4388} {\emph {\bibinfo
  {booktitle} {41st {{AIAA}}/{{ASME}}/{{SAE}}/{{ASEE Joint Propulsion
  Conference}} \&amp; {{Exhibit}}}}}\ (\bibinfo  {publisher} {{American
  Institute of Aeronautics and Astronautics}})\BibitemShut {NoStop}%
\bibitem [{\citenamefont {Ramos-Tomás}\ \emph {et~al.}()\citenamefont
  {Ramos-Tomás}, \citenamefont {Villegas-Prados}, \citenamefont
  {family=Saavedra}, \citenamefont {Cruz},\ and\ \citenamefont
  {Sepúlveda}}]{Ramos-Tomas2024}%
  \BibitemOpen
  \bibfield  {author} {\bibinfo {author} {\bibfnamefont {R.}~\bibnamefont
  {Ramos-Tomás}}, \bibinfo {author} {\bibfnamefont {D.}~\bibnamefont
  {Villegas-Prados}}, \bibinfo {author} {\bibfnamefont {p.~u.}\ \bibnamefont
  {family=Saavedra}, \bibfnamefont {given=Borja}}, \bibinfo {author}
  {\bibfnamefont {J.}~\bibnamefont {Cruz}}, \ and\ \bibinfo {author}
  {\bibfnamefont {B.}~\bibnamefont {Sepúlveda}},\ }\bibfield  {title}
  {\enquote {\bibinfo {title} {Impact of {{Tip Angle}} on the {{Divergence}}
  and {{Efficiency}} of {{Electrospray Thrusters}}},}\ }\href {\doibase
  10.1021/acsaelm.4c01224} {\ \textbf {\bibinfo {volume} {6}},\ \bibinfo
  {pages} {7319--7328}}\BibitemShut {NoStop}%
\bibitem [{\citenamefont {Gamero-Castaño}\ and\ \citenamefont
  {Galobardes-Esteban}()}]{Gamero-Castano2022}%
  \BibitemOpen
  \bibfield  {author} {\bibinfo {author} {\bibfnamefont {M.}~\bibnamefont
  {Gamero-Castaño}}\ and\ \bibinfo {author} {\bibfnamefont {M.}~\bibnamefont
  {Galobardes-Esteban}},\ }\bibfield  {title} {\enquote {\bibinfo {title}
  {Electrospray propulsion: {{Modeling}} of the beams of droplets and ions of
  highly conducting propellants},}\ }\href {\doibase 10.1063/5.0073380} {\
  \textbf {\bibinfo {volume} {131}},\ \bibinfo {pages} {013307}}\BibitemShut
  {NoStop}%
\bibitem [{\citenamefont {Liang}, \citenamefont {Biben},\ and\ \citenamefont
  {Keblinski}()}]{Liang2017}%
  \BibitemOpen
  \bibfield  {author} {\bibinfo {author} {\bibfnamefont {Z.}~\bibnamefont
  {Liang}}, \bibinfo {author} {\bibfnamefont {T.}~\bibnamefont {Biben}}, \ and\
  \bibinfo {author} {\bibfnamefont {P.}~\bibnamefont {Keblinski}},\ }\bibfield
  {title} {\enquote {\bibinfo {title} {Molecular simulation of steady-state
  evaporation and condensation: {{Validity}} of the {{Schrage}}
  relationships},}\ }\href {\doibase 10.1016/j.ijheatmasstransfer.2017.06.025}
  {\ \textbf {\bibinfo {volume} {114}},\ \bibinfo {pages}
  {105--114}}\BibitemShut {NoStop}%
\bibitem [{\citenamefont {Horike}\ \emph {et~al.}()\citenamefont {Horike},
  \citenamefont {Ayano}, \citenamefont {Tsuno}, \citenamefont {Fukushima},
  \citenamefont {Koshiba}, \citenamefont {Misaki},\ and\ \citenamefont
  {Ishida}}]{Horike2018}%
  \BibitemOpen
  \bibfield  {author} {\bibinfo {author} {\bibfnamefont {S.}~\bibnamefont
  {Horike}}, \bibinfo {author} {\bibfnamefont {M.}~\bibnamefont {Ayano}},
  \bibinfo {author} {\bibfnamefont {M.}~\bibnamefont {Tsuno}}, \bibinfo
  {author} {\bibfnamefont {T.}~\bibnamefont {Fukushima}}, \bibinfo {author}
  {\bibfnamefont {Y.}~\bibnamefont {Koshiba}}, \bibinfo {author} {\bibfnamefont
  {M.}~\bibnamefont {Misaki}}, \ and\ \bibinfo {author} {\bibfnamefont
  {K.}~\bibnamefont {Ishida}},\ }\bibfield  {title} {\enquote {\bibinfo {title}
  {Thermodynamics of ionic liquid evaporation under vacuum},}\ }\href {\doibase
  10.1039/C8CP02233J} {\ \textbf {\bibinfo {volume} {20}},\ \bibinfo {pages}
  {21262--21268}}\BibitemShut {NoStop}%
\bibitem [{\citenamefont {Shan}\ \emph {et~al.}()\citenamefont {Shan},
  \citenamefont {Shuai}, \citenamefont {Ma}, \citenamefont {Du}, \citenamefont
  {Dogruoz},\ and\ \citenamefont {Agonafer}}]{Shan2019}%
  \BibitemOpen
  \bibfield  {author} {\bibinfo {author} {\bibfnamefont {L.}~\bibnamefont
  {Shan}}, \bibinfo {author} {\bibfnamefont {S.}~\bibnamefont {Shuai}},
  \bibinfo {author} {\bibfnamefont {B.}~\bibnamefont {Ma}}, \bibinfo {author}
  {\bibfnamefont {Z.}~\bibnamefont {Du}}, \bibinfo {author} {\bibfnamefont
  {B.}~\bibnamefont {Dogruoz}}, \ and\ \bibinfo {author} {\bibfnamefont
  {D.}~\bibnamefont {Agonafer}},\ }\bibfield  {title} {\enquote {\bibinfo
  {title} {Numerical {{Investigation}} of {{Shape Effect}} on {{Microdroplet
  Evaporation}}},}\ }\href {\doibase 10.1115/1.4044962} {\ \textbf {\bibinfo
  {volume} {141}},\ 10.1115/1.4044962}\BibitemShut {NoStop}%
\bibitem [{\citenamefont {MacArthur}, \citenamefont {Colicci},\ and\
  \citenamefont {Lozano}()}]{MacArthur2024}%
  \BibitemOpen
  \bibfield  {author} {\bibinfo {author} {\bibfnamefont {J.}~\bibnamefont
  {MacArthur}}, \bibinfo {author} {\bibfnamefont {V.}~\bibnamefont {Colicci}},
  \ and\ \bibinfo {author} {\bibfnamefont {P.}~\bibnamefont {Lozano}},\
  }\bibfield  {title} {\enquote {\bibinfo {title} {Monodisperse {{Porous
  Emitter Materials}} for {{Ion Electrospray Propulsion}}},}\ }\href {\doibase
  10.2514/1.B39375} {\ \textbf {\bibinfo {volume} {40}},\ \bibinfo {pages}
  {859--868}}\BibitemShut {NoStop}%
\bibitem [{\citenamefont {family=Mora}\ \emph {et~al.}()\citenamefont
  {family=Mora}, \citenamefont {Van~Berkel}, \citenamefont {Enke},
  \citenamefont {Cole}, \citenamefont {Martinez-Sanchez},\ and\ \citenamefont
  {Fenn}}]{Mora2000}%
  \BibitemOpen
  \bibfield  {author} {\bibinfo {author} {\bibfnamefont {p.~l.~u.}\
  \bibnamefont {family=Mora}, \bibfnamefont {given=Juan~Fernandez}}, \bibinfo
  {author} {\bibfnamefont {G.~J.}\ \bibnamefont {Van~Berkel}}, \bibinfo
  {author} {\bibfnamefont {C.~G.}\ \bibnamefont {Enke}}, \bibinfo {author}
  {\bibfnamefont {R.~B.}\ \bibnamefont {Cole}}, \bibinfo {author}
  {\bibfnamefont {M.}~\bibnamefont {Martinez-Sanchez}}, \ and\ \bibinfo
  {author} {\bibfnamefont {J.~B.}\ \bibnamefont {Fenn}},\ }\bibfield  {title}
  {\enquote {\bibinfo {title} {Electrochemical processes in electrospray
  ionization mass spectrometry},}\ }\href {\doibase
  10.1002/1096-9888(200008)35:8<939::AID-JMS36>3.0.CO;2-V} {\ \textbf {\bibinfo
  {volume} {35}},\ \bibinfo {pages} {939--952}}\BibitemShut {NoStop}%
\bibitem [{\citenamefont {Lozano}\ and\ \citenamefont
  {Martínez-Sánchez}({\natexlab{b}})}]{Lozano2004}%
  \BibitemOpen
  \bibfield  {author} {\bibinfo {author} {\bibfnamefont {P.}~\bibnamefont
  {Lozano}}\ and\ \bibinfo {author} {\bibfnamefont {M.}~\bibnamefont
  {Martínez-Sánchez}},\ }\bibfield  {title} {\enquote {\bibinfo {title}
  {Ionic liquid ion sources: Suppression of electrochemical reactions using
  voltage alternation},}\ }\href {\doibase 10.1016/j.jcis.2004.07.037} {\
  \textbf {\bibinfo {volume} {280}},\ \bibinfo {pages} {149--154}
  ({\natexlab{b}})}\BibitemShut {NoStop}%
\bibitem [{\citenamefont {Castro}\ and\ \citenamefont {Fernández de~la
  Mora}()}]{Castro2009}%
  \BibitemOpen
  \bibfield  {author} {\bibinfo {author} {\bibfnamefont {S.}~\bibnamefont
  {Castro}}\ and\ \bibinfo {author} {\bibfnamefont {J.}~\bibnamefont
  {Fernández de~la Mora}},\ }\bibfield  {title} {\enquote {\bibinfo {title}
  {Effect of tip curvature on ionic emissions from {{Taylor}} cones of ionic
  liquids from externally wetted tungsten tips},}\ }\href {\doibase
  10.1063/1.3073873} {\ \textbf {\bibinfo {volume} {105}},\ \bibinfo {pages}
  {034903}}\BibitemShut {NoStop}%
\bibitem [{\citenamefont {Fujiwara}()}]{Fujiwara2020}%
  \BibitemOpen
  \bibfield  {author} {\bibinfo {author} {\bibfnamefont {Y.}~\bibnamefont
  {Fujiwara}},\ }\bibfield  {title} {\enquote {\bibinfo {title}
  {Electrochemical {{Reactions}} of {{Ionic Liquid}} in {{Vacuum}} and {{Their
  Influence}} on {{Ion-Beam Production}} by {{Electrospray}}},}\ }\href
  {\doibase 10.1149/1945-7111/abcb3f} {\ \textbf {\bibinfo {volume} {167}},\
  \bibinfo {pages} {166504}}\BibitemShut {NoStop}%
\bibitem [{\citenamefont {Pieczyńska}\ \emph {et~al.}()\citenamefont
  {Pieczyńska}, \citenamefont {Ofiarska}, \citenamefont {Borzyszkowska},
  \citenamefont {Białk-Bielińska}, \citenamefont {Stepnowski}, \citenamefont
  {Stolte},\ and\ \citenamefont {Siedlecka}}]{Pieczyńska2015}%
  \BibitemOpen
  \bibfield  {author} {\bibinfo {author} {\bibfnamefont {A.}~\bibnamefont
  {Pieczyńska}}, \bibinfo {author} {\bibfnamefont {A.}~\bibnamefont
  {Ofiarska}}, \bibinfo {author} {\bibfnamefont {A.~F.}\ \bibnamefont
  {Borzyszkowska}}, \bibinfo {author} {\bibfnamefont {A.}~\bibnamefont
  {Białk-Bielińska}}, \bibinfo {author} {\bibfnamefont {P.}~\bibnamefont
  {Stepnowski}}, \bibinfo {author} {\bibfnamefont {S.}~\bibnamefont {Stolte}},
  \ and\ \bibinfo {author} {\bibfnamefont {E.~M.}\ \bibnamefont {Siedlecka}},\
  }\bibfield  {title} {\enquote {\bibinfo {title} {A comparative study of
  electrochemical degradation of imidazolium and pyridinium ionic liquids:
  {{A}} reaction pathway and ecotoxicity evaluation},}\ }\href {\doibase
  10.1016/j.seppur.2015.10.045} {\ \textbf {\bibinfo {volume} {156}},\ \bibinfo
  {pages} {522--534}}\BibitemShut {NoStop}%
\bibitem [{\citenamefont {Lozano}()}]{Lozano2003}%
  \BibitemOpen
  \bibfield  {author} {\bibinfo {author} {\bibfnamefont {P.~C.}\ \bibnamefont
  {Lozano}},\ }\bibfield  {title} {\enquote {\bibinfo {title} {Studies on the
  {{Ion-Droplet Mixed Regime}} in {{Colloid Thrusters}}},}\ }\href@noop {} {\
  ,\ \bibinfo {pages} {222}}\BibitemShut {NoStop}%
\bibitem [{\citenamefont {Brikner}\ and\ \citenamefont
  {Lozano}()}]{Brikner2013}%
  \BibitemOpen
  \bibfield  {author} {\bibinfo {author} {\bibfnamefont {N.~A.}\ \bibnamefont
  {Brikner}}\ and\ \bibinfo {author} {\bibfnamefont {P.}~\bibnamefont
  {Lozano}},\ }\bibfield  {title} {\enquote {\bibinfo {title} {Electrostatic
  {{Phenomena}} in {{Ionic Liquid Ion Sources}}},}\ }\href@noop {} {\ ,\
  \bibinfo {pages} {9}}\BibitemShut {NoStop}%
\bibitem [{\citenamefont {Van~Berkel}\ and\ \citenamefont
  {Kertesz}()}]{VanBerkel2001}%
  \BibitemOpen
  \bibfield  {author} {\bibinfo {author} {\bibfnamefont {G.~J.}\ \bibnamefont
  {Van~Berkel}}\ and\ \bibinfo {author} {\bibfnamefont {V.}~\bibnamefont
  {Kertesz}},\ }\bibfield  {title} {\enquote {\bibinfo {title} {Redox buffering
  in an electrospray ion source using a copper capillary emitter},}\ }\href
  {\doibase 10.1002/jms.216} {\ \textbf {\bibinfo {volume} {36}},\ \bibinfo
  {pages} {1125--1132}}\BibitemShut {NoStop}%
\bibitem [{\citenamefont {Heym}\ \emph {et~al.}()\citenamefont {Heym},
  \citenamefont {Korth}, \citenamefont {Thiessen}, \citenamefont {Kern},\ and\
  \citenamefont {Jess}}]{Heym2015}%
  \BibitemOpen
  \bibfield  {author} {\bibinfo {author} {\bibfnamefont {F.}~\bibnamefont
  {Heym}}, \bibinfo {author} {\bibfnamefont {W.}~\bibnamefont {Korth}},
  \bibinfo {author} {\bibfnamefont {J.}~\bibnamefont {Thiessen}}, \bibinfo
  {author} {\bibfnamefont {C.}~\bibnamefont {Kern}}, \ and\ \bibinfo {author}
  {\bibfnamefont {A.}~\bibnamefont {Jess}},\ }\bibfield  {title} {\enquote
  {\bibinfo {title} {Evaporation and {{Decomposition Behavior}} of {{Pure}} and
  {{Supported Ionic Liquids}} under {{Thermal Stress}}},}\ }\href {\doibase
  10.1002/cite.201400139} {\ \textbf {\bibinfo {volume} {87}},\ \bibinfo
  {pages} {791--802}}\BibitemShut {NoStop}%
\end{thebibliography}%
\end{document}